\documentclass[letterpaper,11pt]{article}

\pdfoutput=1

\usepackage{amsmath}
\usepackage{url}
\usepackage{jheppub}
\usepackage{array}
\usepackage{verbatim}
\allowdisplaybreaks

\def\ket#1{\langle #1 \rangle}

\DeclareMathOperator{\Li}{Li}

\newcommand{\be}{\begin{equation}}
\newcommand{\ee}{\end{equation}}
\newcommand{\bea}{\begin{eqnarray}}
\newcommand{\eea}{\end{eqnarray}}
\newcommand{\eqn}[1]{eq.~\eqref{#1}}
\def\fig#1{fig.~{\ref{#1}}}
\def\Fig#1{Fig.~{\ref{#1}}}

\def\eqn#1{eq.~(\ref{#1})}
\def\Eqn#1{Equation~(\ref{#1})}

\def\Eqn#1{Equation~(\ref{#1})}
\def\eqn#1{eq.~(\ref{#1})}

\def\Li{{\rm Li}}

\def\to{\rightarrow}

\def\e{\epsilon}

\newcommand \vev [1] {\langle{#1}\rangle}

\def\beq{\begin{equation}}
\def\eeq{\end{equation}}

\def\bsp#1\esp{\begin{split}#1\end{split}}

\newcommand{\cA}{\begin{cal}A\end{cal}}
\newcommand{\cB}{\begin{cal}B\end{cal}}

\newcommand{\cE}{\begin{cal}E\end{cal}}
\newcommand{\cF}{\begin{cal}F\end{cal}}

\newcommand{\cP}{\begin{cal}P\end{cal}}

\newcommand{\cZ}{\begin{cal}Z\end{cal}}

\newfont{\scyr}{wncyr10 scaled 550}

\def\beq{\begin{equation}}
\def\eeq{\end{equation}}

\def\bsp#1\esp{\begin{split}#1\end{split}}

\title{Heptagons from the Steinmann Cluster Bootstrap}

\author{Lance~J.~Dixon,$^1$}
\author{James~Drummond,$^2$}
\author{Thomas~Harrington,$^3$}
\author{Andrew~J.~McLeod,$^1$}
\author{Georgios~Papathanasiou$^{1,4}$}
\author{and Marcus~Spradlin$^3$}

\affiliation{$^1$ SLAC National Accelerator Laboratory,
Stanford University, Stanford, CA 94309, USA}

\affiliation{$^2$ School of Physics \& Astronomy, University of Southampton \\
Highfield, Southampton, SO17 1BJ, United Kingdom}

\affiliation{$^3$ Department of Physics, Brown University,
Providence, RI 02912, USA}
\affiliation{$^4$ DESY Theory Group, DESY Hamburg, Notkestra{\ss}e 85, D-22607 Hamburg, Germany}

\abstract{We reformulate the heptagon cluster bootstrap to take advantage of the Steinmann relations, which require certain double discontinuities of any amplitude to vanish. These constraints vastly reduce the number of functions needed to bootstrap seven-point amplitudes in planar $\mathcal{N} = 4$ supersymmetric Yang-Mills theory, making higher-loop contributions to these amplitudes more computationally accessible.  In particular, dual superconformal symmetry and well-defined collinear limits suffice to determine uniquely the symbols of the three-loop NMHV and four-loop MHV seven-point amplitudes. We also show that at three loops, relaxing the dual superconformal ($\bar{Q}$) relations and imposing dihedral symmetry (and for NMHV the absence of spurious poles) leaves only a single ambiguity in the heptagon amplitudes.  These results point to a strong tension between the collinear properties of the amplitudes and the Steinmann relations.}

\preprint{
\begin{flushright} Brown--HET--1705\\DESY 16--242 \\ SLAC--PUB--16894
\end{flushright}
}

\begin{document}
\maketitle
\flushbottom


\section{Introduction}
\label{Section:Introduction}

The desire to construct general scattering amplitudes from their analytic and physical properties has been a goal since the birth of the analytic S-matrix program (see e.g.~ref.~\cite{ELOP}).  More recently, such a procedure has been applied in a perturbative context and referred to as bootstrapping.  Aspects of this approach have been applied to theories such as quantum chromodynamics at one loop~\cite{Bern:2005hs,Bern:2005cq,Berger:2006ci} and more recently at two loops~\cite{Dunbar:2016aux,Dunbar:2016cxp,Dunbar:2016gjb}.  However, the most powerful applications to date have been to the planar limit of $\mathcal{N}=4$ super-Yang-Mills (SYM) theory in four dimensions~\cite{Brink:1976bc,Gliozzi:1976qd}.  Fueled by an increased understanding of the classes of analytic functions appearing in amplitudes in general quantum field theories, as well as the stringent constraints obeyed by amplitudes in planar $\mathcal{N}=4$ SYM, it has been possible to advance as far as five loops~\cite{Dixon:2011pw,Dixon:2014iba,Drummond:2014ffa,Dixon:2014xca,Golden:2014pua,Caron-Huot:2016owq}.  These results in turn provide a rich mine of theoretical data for understanding how scattering amplitudes behave.

The planar limit of a large number of colors in $\mathcal{N}=4$ SYM has received a great deal of attention because of the remarkable properties it exhibits. In addition to superconformal symmetry it respects a dual conformal symmetry~\cite{Drummond:2007au,Drummond:2006rz,Bern:2006ew,Bern:2007ct,Alday:2007he}, and amplitudes are dual to polygonal light-like Wilson loops~\cite{Alday:2007hr,Drummond:2007aua,Brandhuber:2007yx,Drummond:2007cf,Drummond:2007au,Bern:2008ap,Drummond:2008aq}. Dual (super)conformal symmetry fixes the four-point and five-point amplitudes uniquely to match the Bern-Dixon-Smirnov (BDS) ansatz~\cite{Bern:2005iz}, which captures all the infrared divergences of planar scattering amplitudes. Starting at six points, the BDS ansatz receives corrections from finite functions of dual conformal invariants~\cite{Drummond:2007bm,Bartels:2008ce,Drummond:2008aq,Bern:2008ap}. The correction to the maximally helicity violating (MHV) amplitude has traditionally been expressed in terms of a (BDS) remainder function~\cite{Bern:2008ap,Drummond:2008aq,Dixon:2011pw,Dixon:2013eka,Drummond:2014ffa}, while the correction to the next-to-maximally helicity violating (NMHV) amplitude has traditionally been expressed in terms of the infrared-finite NMHV ratio function~\cite{Drummond:2008vq,Drummond:2008bq,Kosower:2010yk,Dixon:2011nj,Dixon:2014iba,Dixon:2015iva}.

The cluster bootstrap program is built on the idea that certain scattering amplitudes can be determined order by order in perturbation theory using a set of basic building blocks known as cluster coordinates~\cite{FG03b,GSV}. Inspired by the results of refs.~\cite{Goncharov:2010jf,Golden:2013xva}, the bootstrap approach developed in refs.~\cite{Dixon:2011pw,Dixon:2014iba,Drummond:2014ffa,Dixon:2014xca,Golden:2014pua,Caron-Huot:2016owq} assumes that the MHV and NMHV amplitudes at each loop order belong to a particular class of iterated integrals, or generalized polylogarithms. More specifically, the $L$-loop contribution to the remainder and ratio functions is expected to lie within the space spanned by polylogarithms of weight $2L$~\cite{ArkaniHamed:2012nw} whose symbols can be written in terms of cluster $\mathcal{A}$-coordinates. A further constraint on the relevant space of functions comes from the restriction that only physical branch cuts can appear in the remainder and ratio functions~\cite{Gaiotto:2011dt}.

To make use of this expectation, in the bootstrap program one first constructs a general linear combination of the above set of functions to serve as an ansatz.  Then one tries to determine all free coefficients in the ansatz by imposing analytic and physical constraints. This procedure becomes increasingly computationally expensive at higher loop orders, largely due to the fact that the number of relevant functions increases exponentially with the weight. It is hoped that one day a constructive procedure for determining these amplitudes can be developed that does not require constructing the full weight-$2L$ space as an intermediate step.  A promising candidate in this respect is the Wilson loop Operator Product Expansion (OPE)~\cite{Alday:2010ku,Gaiotto:2011dt,Sever:2011pc} and the Pentagon OPE program~\cite{Basso:2013vsa,Basso:2013aha,Basso:2014nra,Belitsky:2015efa,Basso:2015rta,Basso:2015uxa,Belitsky:2016vyq} which provides finite-coupling expressions for the amplitudes as an expansion around (multi-)collinear kinematics.  The main challenge in this framework is to resum the infinite series around these kinematics; there has been progress recently in this direction at weak coupling~\cite{Drummond:2015jea,Cordova:2016woh,Lam:2016rel}.  Another potential constructive approach could involve the Amplituhedron~\cite{Arkani-Hamed:2013jha,Arkani-Hamed:2013kca} description of the multi-loop integrand. Perhaps one can extend the methods of ref.~\cite{DPSSV} for reading off the branch-point locations, in order to enable reading off the entire function.

To date, six- and seven-point amplitudes have been computed in the cluster bootstrap program through the study of so-called hexagon and heptagon functions. Both helicity configurations of the six-point amplitude have been determined through five loops~\cite{Caron-Huot:2016owq}, while the MHV seven-point amplitude has been determined at symbol level through three loops~\cite{Drummond:2014ffa}. The seven-point NMHV amplitude has not yet received attention in the bootstrap program, but it has been calculated through two loops using slightly different methods~\cite{CaronHuot:2011kk}. Surprisingly, bootstrapping the seven-point remainder function has thus far proven to be conceptually simpler (i.e.~requiring the imposition of fewer constraints) than bootstrapping its six-point counterpart. The collinear limit of the seven-point remainder function must be nonsingular and a well-defined hexagon function. This requirement is so restrictive that it entirely determines the two-loop heptagon remainder function, up to an overall scale. It similarly determines the three-loop remainder function, once the full implications of dual superconformal symmetry are taken into account~\cite{Drummond:2014ffa}. The corresponding hexagon remainder function symbols may then be obtained by taking a collinear limit.

In a recent breakthrough~\cite{Caron-Huot:2016owq}, the classic work of Steinmann~\cite{Steinmann,Steinmann2} on the compatibility of branch cuts in different channels has been used to supercharge the hexagon function bootstrap program. The Steinmann relations dramatically reduce the size of the functional haystack one must search through in order to find amplitudes, putting higher-loop amplitudes that were previously inaccessible within reach. In this paper we reformulate the heptagon bootstrap of ref.~\cite{Drummond:2014ffa} to exploit the power of the Steinmann relations. With their help, we are able to fully determine the symbol of the seven-point three-loop NMHV and four-loop MHV amplitude in planar $\mathcal{N}=4$ SYM, using only a few simple physical and mathematical inputs.  In a separate paper~\cite{DDHMPS2}, we will investigate various kinematical limits of these amplitudes in more detail, including the multi-Regge limit~\cite{Bartels:2008ce,Bartels:2008sc,Lipatov:2010ad,Bartels:2011ge,Fadin:2011we,Lipatov:2012gk,Dixon:2012yy,Bartels:2013jna,Bartels:2014jya,Bargheer:2016eyp,Broedel:2016kls,DelDuca:2016lad}, the OPE limit~\cite{Alday:2010ku,Gaiotto:2011dt,Sever:2011pc,Basso:2013vsa,Basso:2013aha,Basso:2014nra,Belitsky:2015efa,Basso:2015rta}, and the self-crossing limit~\cite{Georgiou:2009mp,Dixon:2016epj}.  In this paper, we study one of the simpler limits, where the NMHV seven-point amplitude factorizes on a multi-particle pole.

This paper is organized as follows. In section~\ref{Section:SevenParticleScatteringAmplitudes} we begin by reviewing the general structure of seven-particle MHV and NMHV (super)amplitudes, and different schemes for subtracting their infrared divergences. Section~\ref{Section:BootstrapToolbox} discusses the essential ingredients of the amplitude bootstrap for constructing heptagon functions, which are believed to describe the nontrivial kinematical dependence of these amplitudes.  Section~\ref{Section:MHVandNMHVConstraints} focuses on the additional physical constraints that allow us to single out the MHV or NMHV amplitude from this space of functions.

Our main results, including the analysis of the general space of heptagon symbols, and the determination of the three-loop NMHV and four-loop MHV amplitude symbols, are presented in section~\ref{Section:Results}. Section~\ref{Section:Multifact} describes a sample kinematical limit, the behavior of the NMHV amplitude as a multi-particle Mandelstam invariant vanishes.  Finally, section~\ref{Section:Discussion} contains our conclusions, and discusses possible avenues for future study. 

Many of the analytic results in this paper are too lengthy to present in the manuscript.  Instead, computer-readable files containing our results can be downloaded from~\cite{HeptagonsPage}.


\section{Seven-Particle Scattering Amplitudes}
\label{Section:SevenParticleScatteringAmplitudes}

\subsection{MHV: The Remainder Function}
\label{Section:RemainderFunction}

In planar $\mathcal{N}=4$ SYM, $n$-particle amplitudes are completely characterized by the color-ordered partial amplitudes $A_n$, which are the coefficients of specific traces ${\rm Tr}(T^{a_1}T^{a_2}\cdots T^{a_n})$ in the color decomposition of the amplitudes.  The MHV helicity configuration has precisely two gluons with negative helicity and $(n{-}2)$ with positive helicity (in a convention where all particles are outgoing).  The MHV amplitude is encoded in the remainder function $R_n$, which is defined by factoring out the BDS ansatz $A_{n}^{\text{BDS}}$~\cite{Bern:2005iz} (reviewed in appendix~\ref{Appendix:BDSAnsatz}):
\begin{equation}
\label{eq:remainderdefinition}
A_{n}^{\text{MHV}} = A_{n}^{\text{BDS}}\ \text{exp}\left[R_{n}\right].
\end{equation}
The BDS ansatz captures all the infrared and collinear divergences~\cite{Magnea:1990zb,Catani:1998bh,Sterman:2002qn} in the planar amplitude, so the
remainder function is infrared finite.  It is also invariant under dual conformal transformations~\cite{Drummond:2006rz,Bern:2006ew,Alday:2007hr,Bern:2007ct,Drummond:2007au}.  Moreover, since the BDS ansatz accounts for collinear factorization to all orders in perturbation theory~\cite{Bern:2005iz}, the $n$-point remainder function smoothly tends to the $(n{-}1)$-point remainder function in its collinear limits, a fact that will prove to be an important ingredient in the bootstrap program.

In the definition~\eqref{eq:remainderdefinition}, $R_{n}$ is the finite-coupling (or all-loop) remainder function. Here we will be interested in its perturbative expansion. For any function $F$ of the coupling, we denote the coefficients of its perturbative expansion with a superscript according to the definition
\be\label{eq:gLoopExpansion}
F=\sum_{L=0}^\infty g^{2L} F^{(L)}\,,
\ee
where $g^2 = {g_{YM}^{2}N}/{(16 \pi^{2})}$, $g_{YM}$ is the Yang-Mills coupling constant, and $N$ is the number of colors.  Elsewhere in the literature, the coupling constant $a=2g^2$ is often used. The $L$-loop contribution to the remainder function, $R_n^{(L)}$, is expected to be a weight-$2L$ iterated integral.

The remainder function vanishes for the four- and five-particle amplitudes, because dual conformally invariant cross ratios cannot be formed with fewer than six external lightlike momenta (in other words, the BDS ansatz is correct to all loop orders for $n=4$ or 5)~\cite{Drummond:2007bm,Drummond:2008aq,Bern:2008ap}. The first nontrivial case, the six-point remainder function, has been successfully computed at two loops~\cite{Goncharov:2010jf}, three loops~\cite{Dixon:2011pw,CaronHuot:2011kk,Dixon:2013eka}, four loops~\cite{Dixon:2014voa} and recently five loops~\cite{Caron-Huot:2016owq}. At seven points, the remainder function has been computed at two loops~\cite{CaronHuot:2011ky,CaronHuot:2011kk,Golden:2013lha,Golden:2014xqf} and its symbol has been computed at three loops~\cite{Drummond:2014ffa}. The symbol of the four-loop seven-point MHV remainder function $R_7^{(4)}$ is one of the main results of this paper.

\subsection{NMHV: The Ratio Function and $R$-invariants}
Beyond the MHV case, scattering amplitudes in SYM theory are most efficiently organized by exploiting the (dual) superconformal symmetry~\cite{Drummond:2008vq} of the theory, as reviewed in ref.~\cite{Drummond:2010km}.

In a nutshell, one starts by packaging the on-shell particle content of the theory into a single superfield $\Phi$ with the help of four Grassmann variables $\eta^A$, whose index transforms in the fundamental representation of the $SU(4)$ $R$-symmetry group. In other words, all external states, gluons $G^\pm$, fermions $\Gamma_A$ and $\bar \Gamma^A$, and scalars $S_{AB}$, can be simultaneously described by the superfield
\be
\Phi=G^+ +\eta^A\Gamma_A+\tfrac{1}{2!}\eta^A\eta^B S_{AB}+\tfrac{1}{3!}\eta^A\eta^B\eta^C\epsilon_{ABCD}\bar \Gamma^D+\tfrac{1}{4!}\eta^A\eta^B\eta^C\eta^D\epsilon_{ABCD}G^-\,,
\ee
which allows us to combine all $n$-point amplitudes into a superamplitude $\mathcal{A}_n(\Phi_1,\ldots,\Phi_n)$.

Expanding the superamplitude in the Grassmann variables separates out its different helicity components. The MHV amplitude is contained in the part of $\mathcal{A}_n^\text{MHV}$ with 8 powers of Grassmann variables, or Grassmann degree 8. Specifically, the MHV amplitude discussed in the previous subsection is given in the MHV superamplitude by the term
\begin{align}\label{gen-MHV}
{\cal A}_{n}^{\rm MHV} = (2\pi)^4 \delta^{(4)} \Bigl(\sum_{i=1}^n p_i\Bigr)
\sum_{1\le j<k\le n}
(\eta_j)^4 (\eta_k)^4 A^{\rm MHV} _n(1^+...\, j^-...\, k^-...\, n^+) + \ldots,
\end{align}
where we have shown only the pure-gluon terms explicitly.
Similarly, the terms of Grassmann degree $12$ make up the NMHV superamplitude. Since NMHV amplitudes in this theory have the same infrared-divergent structure as MHV amplitudes, the two superamplitudes can be related by
\be\label{eq:ratiofn}
\mathcal{A}^{\text{NMHV}}_n=\mathcal{A}^{\text{MHV}}_n \, \cP_n\,,
\ee
where the infrared-finite quantity $\cP_n$ is called the NMHV ratio function and has Grassmann degree 4. On the basis of tree-level and one-loop amplitude computations, it was argued in ref.~\cite{Drummond:2008vq} that $\cP_n$ is dual conformally invariant.

At tree level, the dual conformal symmetry is enhanced to dual superconformal symmetry, and the ratio function can be written as a sum of \emph{dual superconformal invariants} or `\emph{$R$-invariants}'~\cite{Drummond:2008vq,Drummond:2008bq}. These quantities, which carry the dependence on the fermionic variables, are algebraic functions of the kinematics and can be written as Grassmannian contour integrals~\cite{Mason:2009qx}. From this representation it is also possible to prove their invariance under ordinary superconformal transformations~\cite{ArkaniHamed:2009dn,ArkaniHamed:2009vw}, or in other words their Yangian invariance~\cite{Drummond:2010qh}.

As shown in ref.~\cite{Mason:2009qx}, $R$-invariants are most easily expressed in terms of the momentum supertwistors $\mathcal{Z}_i$ defined by\footnote{The indices $\alpha,\dot \alpha=1,2$ denote the components of the spinor representation of the Lorentz group $SO(3,1)\simeq SL(2,\mathbb{C})$.}~\cite{Hodges:2009hk}
\be\label{eq:supertwistors}
\mathcal{Z}_i = (Z_i \, | \, \chi_i)\,, \qquad
Z_i^{\alpha,\dot\alpha} =
(\lambda_i^\alpha , x_i^{\beta \dot\alpha}\lambda_{i\beta})\,,
\qquad
\chi_i^A= \theta_i^{\alpha A}\lambda_{i \alpha} \,.
\ee
Their fermionic components $\chi_i$ are associated with the fermionic dual coordinates $\theta_i$ in the same way that the bosonic twistors $Z_i$ are associated with the bosonic dual coordinates $x_i$. Differences between color-adjacent dual coordinates $x_i$ and $\theta_i$ are related to the external momenta $p_i$ and supermomenta $q_i$, respectively:
\be
p_i^{\alpha \dot\alpha} =
\lambda_i^\alpha \tilde{\lambda}_i^{\dot\alpha}
= x_{i+1}^{\alpha \dot\alpha} - x_{i}^{\alpha \dot\alpha}\,, \qquad
q_i^{\alpha A} = \lambda_i^{\alpha} \eta_i^A
= \theta_{i+1}^{\alpha A} - \theta_{i}^{\alpha A}\,.
\label{xthetadef}
\ee
Given any set of five supertwistors $\cZ_a, \cZ_b, \cZ_c, \cZ_d, \cZ_e$, we may define a corresponding NMHV $R$-invariant as a 5-bracket
\be\label{fivebrak}
[abcde] = \frac{\delta^{0|4} \big( \chi_a \langle b c d e \rangle + \text{cyclic} \big)}{\langle a b c d \rangle \langle b c d e \rangle \langle c d e a \rangle \langle d e a b \rangle \langle e a b c \rangle} \,,
\ee
in terms of dual conformally invariant bosonic 4-brackets
\be\label{fourbrak}
\vev{ijkl} \equiv\langle Z_{i} Z_j Z_{k} Z_l \rangle=\epsilon_{ABCD} Z_{i}^{A}Z_{j}^{B}Z_{k}^{C}Z_{l}^{D}=\det(Z_i Z_j Z_k Z_l)\,,
\ee
and a fermionic delta function $\delta^{0|4}(\xi)=\xi^1 \xi^2 \xi^3 \xi^4$ for the different $SU(4)$ components of $\xi$. The original definition of the $R$-invariants~\cite{Drummond:2008vq,Drummond:2008bq} (there denoted $R_{r;ab}$) in normal twistor space corresponds to the special case $R_{r;ab} = [r, \, a{-}1, \, a, \, b{-}1, \, b]$.

From the definition~\eqref{fivebrak}, we can see that $R$-invariants are antisymmetric in the exchange of any pair of supertwistor indices (hence also invariant under cyclic permutations). They are also manifestly dual conformally invariant, since they don't depend on spinor products $\langle ij\rangle$. The aforementioned Grassmannian contour integral representation in momentum twistor space~\cite{Mason:2009qx} makes the full dual conformal invariance manifest. It also allows one to prove more transparently the following important identity between $R$-invariants: Given any six momentum supertwistors $\cZ_a, \cZ_b, \cZ_c, \cZ_d, \cZ_e, \cZ_f$, their $R$-invariants are related by~\cite{Drummond:2008vq}
\be\label{sixZidentity}
[abcde] - [bcdef] + [cdefa] - [defab] + [efabc] - [fabcd] = 0\,.
\ee
For $n$-particle scattering, there exist $\binom{n}{6}$ such equations for the $\binom{n}{5}$ distinct $R$-invariants; however, it turns out that only $\binom{n-1}{5}$ are independent. So in the end we are left with
\be
\#\,\,\text{linearly independent}\,\,n\text{-particle}\,\,R\text{-invariants}=\binom{n}{5}-\binom{n-1}{5}=\binom{n-1}{4}\,.
\ee
For example, there are 5, 15, and 35 independent $R$-invariants relevant for 6-, 7- and 8-particle NMHV scattering amplitudes, respectively.

Let us now focus on the seven-particle NMHV superamplitude. For compactness we may express the corresponding $R$-invariants in terms of the particle indices that are {\it not} present in the 5-brackets~\eqref{fivebrak}, for example
\be
[12345]=(67)=(76)\,,
\ee
where (by convention) the 5-bracket on the left-hand side of this definition is always ordered, so ordering on the right-hand side doesn't matter.

In this notation, the representation for the tree-level ratio function found in ref.~\cite{Drummond:2008bq} may be rewritten as
\be\label{eq:P7tree}
\cP^{(0)}_7 = \frac{3}{7} \, (12) + \frac{1}{7} \, (13) + \frac{2}{7} \, (14) ~ + \text{cyclic}\,.
\ee
Following the same reference, we find it convenient to use a basis of 15 independent $R$-invariants consisting of $\cP^{(0)}_7$ together with $(12),(14)$, and their cyclic permutations. (Because $\cP^{(0)}_7$ is totally symmetric, it has no independent cyclic images.) In particular, the remaining $R$-invariants $(i,i+2)$ are related to this set by
\be\label{eq:DependentRInvariant}
 {(13)} = - \, {(15)}-{(17)}-{(34)}-{(36)}-{(56)}+\cP^{(0)}_7\,,
\ee
plus the cyclic permutations of this identity.

Beyond tree level, the independent $R$-invariants are dressed by transcendental functions of dual conformal invariants, and the ratio function can be put in the form
\be\label{eq:P7}
\cP_7 = \cP^{(0)}_7 \, V_0 + \big[ (12) \, V_{12} + (14) \, V_{14} ~ + \text{cyclic} \big]\,.
\ee
As we will review in section~\ref{Section:DiscreteSymmetries}, $\cP_7$ is symmetric under the dihedral group $D_7$.  The component $V_0$ inherits the full dihedral symmetry of $\cP^{(0)}_7$, whereas $V_{12}$ and $V_{14}$ are only invariant under the flip $i\to 3{-}i$ and $i\to 5{-}i$ of their momentum twistor labels, respectively.

The dependence of $\cP_7$ on the coupling enters only through the functions $V_0$ and $V_{ij}$. Their $L$-loop contributions, $V_0^{(L)}$ and $V_{ij}^{(L)}$, like the remainder function, $R_7^{(L)}$, are expected to be weight-$2L$ iterated integrals.  Using the notation introduced in~\eqn{eq:gLoopExpansion} we must have
\be
V^{(0)}_0=1\,,\quad V^{(0)}_{12}=V^{(0)}_{14}=0\,
\ee
at tree level. At one loop, these functions become~\cite{Drummond:2008bq}
\be
\begin{aligned}
V^{(1)}_0&=\Li_2\left(1-u_1\right)-\Li_2\left(1-u_1 u_4\right)-\log u_1 \log u_3 + \text{ cyclic}\,,\\
V^{(1)}_{12}&=-\Li_2\left(1-u_6\right)+\Li_2\left(1-u_1 u_4\right)+\Li_2\left(1-u_2 u_6\right)+\Li_2\left(1-u_3 u_6\right),\\
&+\log u_1 \log u_2-\log u_3 \log u_2+\log u_4 \log u_2+\log u_1 \log u_3+\log u_3 \log u_4\\
&+\log u_1 \log u_6+\log u_4 \log u_6-\zeta_2\,,\\
V^{(1)}_{14}&=\Li_2\left(1-u_1 u_4\right)+\Li_2\left(1-u_3 u_6\right)+\log u_1 \log u_3+\log u_4 \log u_3+\log u_1 \log u_6\\
&+\log u_4 \log u_6-\zeta_2\,.
\end{aligned}
\ee
See also ref.~\cite{ArkaniHamed:2010gh} for a more recent, compact representation of the same amplitude. In the above relations and everything that follows, the cross ratios $u_i$ are defined by,
\be
u_{ij}=\frac{x^2_{i,j+1} \, x^2_{i+1,j}}{x^2_{i,j} \, x^2_{i+1,j+1}} \,,\qquad\quad
u_{i}=u_{i+1,i+4}=\frac{x^2_{i+1,i+5} \, x^2_{i+2,i+4}}
                    {x^2_{i+1,i+4} \, x^2_{i+2,i+5}} \,.
\label{uidef}
\ee
The $u_i$ are dual conformally invariant combinations of the Mandelstam invariants, see~\eqn{xthetadef} and also~\eqn{x2p} below. 

Finally, the symbol of the two-loop NMHV heptagon has been computed in ref.~\cite{CaronHuot:2011kk} using the same choice of independent $R$-invariants as in~\eqn{eq:P7}, with the help of an anomaly equation for the $\bar{Q}$ dual superconformal symmetry generators.  Here we will use the Steinmann cluster bootstrap to push to three loops:  The symbols of the functions $V_0^{(3)}$, $V_{12}^{(3)}$, and $V_{14}^{(3)}$ constituting the three-loop seven-point NMHV ratio function are another of the main results of this paper.

\subsection{The BDS- and BDS-like Normalized Amplitudes}
\label{Section:BDSVersusBDSlike}

In the previous sections we mentioned that MHV and NMHV amplitudes have the same infrared-divergent structure, which is accurately captured by the BDS ansatz. This fact allows us to define the MHV and NMHV \emph{BDS-normalized} superamplitudes,
\begin{align}
\cB_n\equiv\frac{\cA_n^{\text{MHV}}}{\cA_n^{\text{BDS}}}&= \frac{A_n^{\text{MHV}}}{A_n^{\text{BDS}}}=\exp\left[R_n\right], \label{eq1} \\
B_n\equiv\frac{\cA_n^{\text{NMHV}}}{\cA_n^{\text{BDS}}}&=\frac{\cA_n^{\text{NMHV}}}{\cA_n^{\text{MHV}}}\frac{\cA_n^{\text{MHV}}}{\cA_n^{\text{BDS}}}=\cP_n \, \cB_n\label{eq2} \,,
\end{align}
where $\cA_n^{\text{BDS}}$ is the superamplitude obtained from the bosonic BDS ansatz by replacing the tree-level MHV Parke-Taylor factor~\cite{Parke:1986gb,Berends:1987me} it contains with its supersymmetrized version~\cite{Nair:1988bq}. Indeed, normalizations~\eqref{eq1}, \eqref{eq2} were found to be more natural for the study of the dual superconformal symmetry anomaly equation~\cite{CaronHuot:2011kk}.

In what follows, it will prove greatly beneficial to define yet another set of infrared-finite quantities, using an alternate normalization factor that is compatible with the Steinmann relations. The BDS ansatz is essentially the exponential of the full one-loop amplitude, which includes a finite part with nontrivial dependence on Mandelstam invariants involving all possible numbers of external momenta.  Dividing by the BDS ansatz produces a quantity with altered dependence on three-particle Mandelstam invariants.  As we will see, such a quantity does not satisfy the Steinmann relations. In the case of seven-particle scattering (indeed, whenever $n$ is not a multiple of four), all the dependence on the three-particle invariants (and higher-particle invariants) can be assembled into a dual conformally invariant function $Y_n$, which we may remove from the one-loop amplitude in order to define a \emph{BDS-like} ansatz,
\be
\cA_n^{\text{BDS-like}}\equiv\cA_n^{\text{BDS}}\exp\left[\frac{\Gamma_{\text{cusp}}}{4} Y_n\right],
\label{BDSlikeBDSn}
\ee
where 
\bea
Y_6 &=& - \, \Li_2\left(1-\frac{1}{u}\right)
         -\Li_2\left(1-\frac{1}{v}\right)
         -\Li_2\left(1-\frac{1}{w}\right)\,,\label{Y6def}\\
Y_7 &=& -\sum_{i=1}^7 \biggl[ \Li_2\left(1-\frac{1}{u_i}\right)
  + \frac{1}{2} \log \left(\frac{u_{i+2}u_{i{-}2}}{u_{i+3}u_{i}u_{i{-}3}}\right)
               \log u_i \biggr]\,,
\label{Y7def}
\eea
and
\be\label{gamma_cusp}
\Gamma_{\text{cusp}}=\sum_{L=1}^\infty g^{2L}\Gamma_{\text{cusp}}^L= 4g^2-\frac{4\pi ^2}{3} g^4 +\frac{44 \pi ^4}{45}g^6- 4\left(\frac{73 \pi ^6}{315}+8 \zeta_3^2\right)g^8+\mathcal{O}(g^{10})\,,
\ee
is the cusp anomalous dimension in the normalization of e.g.~\cite{Basso:2013aha}.\footnote{In particular, $\Gamma_{\text{cusp}} = \gamma_K/2$ compared to the normalization of~\cite{Bern:2005iz} and subsequent papers of Dixon and collaborators.} In \eqn{Y6def}, $u,v,w$ are the three cross ratios for six-point kinematics, defined below in \eqn{uvw_def}. The difference between the BDS- and BDS-like-normalized ans\"atze for seven-point kinematics is reviewed in more detail in appendix~\ref{Appendix:BDSAnsatz}. The utility of the BDS-like ansatz was first noticed in the strong coupling analysis of amplitudes via the AdS/CFT correspondence~\cite{Alday:2009dv} (see also~ref.~\cite{Yang:2010as}).  At weak coupling, it was found to simplify the six-point multi-particle factorization limit~\cite{Dixon:2014iba}, self-crossing limit~\cite{Dixon:2016epj} and NMHV $\bar{Q}$ relations~\cite{Dixon:2015iva}, before its role in applying the six-point Steinmann relations was noticed~\cite{Caron-Huot:2016owq}.  We will see its advantages as well in our seven-point analysis.

When $n$ is a multiple of four it is not possible to simultaneously remove the dependence on all three-particle and higher-particle Mandelstam invariants in a conformally invariant fashion~\cite{Yang:2010az}. However, for $n=8$ it is still possible to separately remove the dependence of all three-particle invariants, \emph{or} of all four-particle invariants, giving rise to two different BDS-like ans\"atze.

Restricting our attention to the case $n\nmid 4$, we may thus define the \emph{BDS-like-normalized} MHV and NMHV amplitudes as
\be\label{eq:BDS-like_def}
\begin{aligned}
\cE_n&\equiv\frac{\cA_n^{\text{MHV}}}{\cA_n^{\text{BDS-like}}}=\frac{\cA_n^{\text{MHV}}}{\cA_n^{\text{BDS}}}\frac{\cA_n^{\text{BDS}}}{\cA_n^{\text{BDS-like}}}=\cB_n\exp\left[-\frac{\Gamma_{\text{cusp}}}{4} Y_n\right]=\exp\left[R_n-\frac{\Gamma_{\text{cusp}}}{4} Y_n\right],\\
E_n&\equiv\frac{\cA_n^{\text{NMHV}}}{\cA_n^{\text{BDS-like}}}=\frac{\cA_n^{\text{NMHV}}}{\cA_n^{\text{BDS}}}\frac{\cA_n^{\text{BDS}}}{\cA_n^{\text{BDS-like}}}=B_n\exp\left[-\frac{\Gamma_{\text{cusp}}}{4} Y_n\right]=\cP_n \, \cE_n\,,
\end{aligned}
\ee
where we have also spelled out their relation to the previously-considered normalizations. Note that
\be
\cE^{(1)}_n=-Y_n\,,
\ee
since $R_n$ starts at two loops.

Because we will focus almost exclusively on heptagon amplitudes in this paper, we will usually drop the particle index $n$ from of all of its associated quantities in order to avoid clutter, e.g.~$\cP_7\to \cP$, $\cE_7\to \cE$ and $E_7\to E$. In the NMHV case we will instead use subscripts to denote components multiplying the different $R$-invariants. For example, the BDS-normalized and BDS-like-normalized analogs of~\eqn{eq:P7} are
\begin{align}
B &= \cP^{(0)} \, B_0 + \big[ (12) \, B_{12} + (14) \, B_{14} + \text{ cyclic} \big]\,,
\label{eq:B7components}\\
E &= \cP^{(0)} \, E_0 + \big[ (12) \, E_{12} + (14) \, E_{14} + \text{ cyclic} \big]\,.
\label{eq:E7components}
\end{align}
It is important to note that because the $R$-invariants are coupling-independent, the same coupling-dependent factor that relates NMHV superamplitudes in different normalizations will also relate the respective coefficient functions of the $R$-invariants. In other words,
\be\label{eq:E7componentBDSToBDSlike}
E_{*} = B_{*}\exp\left[-\frac{\Gamma_{\text{cusp}}}{4} Y\right] = \cE \, V_{*}\,,
\ee
where $*$ can be any index, 0 or $ij$.

Given that in this paper we will be focusing exclusively on symbols, it's also worth emphasizing that when expanding~\eqn{eq:BDS-like_def} or equivalently~\eqn{eq:E7componentBDSToBDSlike} at weak coupling, we may replace $\Gamma_{\text{cusp}}\to4g^2$, as a consequence of the fact that the symbol of any term containing a transcendental constant, such as $\zeta_n$, is zero. Thus, the conversion between the BDS-like-normalized quantities $F\in\{\cE,E,E_0,E_{ij}\}$ and the corresponding BDS-normalized quantities $\cF\in\{\cB,B,B_0,B_{ij}\}$ at symbol level and at fixed order in the coupling, simply becomes
\be\label{eq:BDSLikeToBDS}
  F^{(L)}=\sum_{k=0}^L \cF^{(k)} \frac{(-Y_n)^{L-k}}{(L-k)!} \,, \qquad\quad
  \cF^{(L)}=\sum_{k=0}^L F^{(k)} \frac{Y_n^{L-k}}{(L-k)!}\,.
\ee
In particular, for $R_7$, which sits in the exponent, its analogous conversion to $\mathcal{E}_7$ through four loops is given by
\begin{align}
\mathcal{E}_{7}^{(2)} &= R_{7}^{(2)} +\frac{1}{2}{\left(\mathcal{E}_{7}^{(1)}\right)^{2}}\,, \nonumber \\
\label{eq:ERrelations}
\mathcal{E}_{7}^{(3)} &= R_{7}^{(3)} +\mathcal{E}_{7}^{(1)}R_{7}^{(2)} +\frac{1}{6}{\left(\mathcal{E}_{7}^{(1)}\right)^{3}}\,, \\
\mathcal{E}_{7}^{(4)} &= R_{7}^{(4)} + \frac{1}{2}{\left(R_{7}^{(2)}\right)^{2}}
+ {\mathcal{E}_{7}^{(1)}R_{7}^{(3)}} 
+ \frac{1}{2}{\left(\mathcal{E}_{7}^{(1)}\right)^{2}R_{7}^{(2)}}
+ \frac{1}{24}{\left(\mathcal{E}_{7}^{(1)}\right)^{4}}\,. \nonumber
\end{align}

In summary, all the nontrivial kinematic dependence of seven-particle scattering can be encoded in the four transcendental functions $R_7, B_0, B_{12}$ and $B_{14}$ using BDS normalization, or equivalently $\cE, E_0, E_{12}$ and $E_{14}$ using BDS-like normalization. (The other $E_{ij}$ that are needed are related to $E_{12}$ and $E_{14}$ by cyclic permutations.) These functions are all expected to belong to a very special class of transcendental functions called heptagon functions, whose definition and construction we turn to in the next section. However, we will see that it is only the BDS-like-normalized amplitudes that inherit a specific analytic property from the full amplitudes: they satisfy the Steinmann relations. Taking this restriction into account hugely trims the space of heptagon functions needed to bootstrap the BDS-like normalized functions, thus allowing for a far more efficient construction of the amplitude.


\section{The Steinmann Cluster Bootstrap}
\label{Section:BootstrapToolbox}

The heptagon bootstrap approach we use in this paper is a slight refinement of that used in ref.~\cite{Drummond:2014ffa}, which in turn is a generalization of the hexagon function bootstrap~\cite{Dixon:2011pw, Dixon:2011nj, Dixon:2013eka, Dixon:2014voa, Dixon:2014iba, Dixon:2014xca}. We begin this section by reviewing some basics of the bootstrap approach and defining heptagon functions. Then we express the seven-point Steinmann relations in the language of cluster $\mathcal{A}$-coordinates. We assume a basic working knowledge of both symbols~\cite{Goncharov:2010jf, Gonch3, Gonch2, FBThesis, Gonch, Duhr:2011zq, Golden:2013xva, Golden:2014xqa} and momentum twistor notation~\cite{Hodges:2009hk}.

\subsection{Symbol Alphabet}
\label{Section:SymbolAlphabet}

In the cluster bootstrap program for $n$-point amplitudes in planar SYM theory, we assume that the symbol alphabet consists of certain objects known as cluster $\mathcal{A}$-coordinates. These coordinates have been discussed extensively in the context of scattering amplitudes; see for example ref.~\cite{Golden:2013xva}. Here we will only briefly recall that the kinematic data for a scattering process in planar SYM theory may be specified by a collection of $n$ momentum twistors~\cite{Hodges:2009hk}, each of which is a homogeneous coordinate $Z_i$ on $\mathbb{P}^3$. The configuration space for SYM theory is $\text{Conf}_{n} (\mathbb{P}^{3}) = \text{Gr}(4,n)/(\mathbb{C}^{*})^{n-1}$, and cluster $\mathcal{A}$-coordinates on this space can be expressed in terms of the Pl\"ucker coordinates of 4-brackets $\ket{ijkl}$, which we defined in~\eqn{fourbrak}.

Mandelstam invariants constructed from sums of cyclically adjacent external momenta $p_i, p_{i+1},\ldots,p_{j-1}$ can be expressed nicely in terms of dual coordinates $x_i$ satisfying the relation $p_i = x_{i+1} - x_i$. Using the notation $x_{ij} = x_{i}-x_{j}$, the Mandelstam invariant $s_{i,\dots,j-1}$ can be written as
\begin{equation}
\label{x2p}
s_{i,\dots,j-1} = (p_{i} + p_{i+1} + \dots + p_{j-1})^{2} = x_{ij}^{2} =
\frac{\ket{i{-}1\, i\, j{-}1\, j}}{\ket{i{-}1\, i}\ket{j{-}1\, j}}\,.
\end{equation}
Here we have also shown how to express the Mandelstam invariant $s_{i,\ldots,j-1}$ in terms of Pl\"ucker coordinates and the usual spinor products $\ket{ij}=\epsilon_{\alpha\beta}\lambda_i^\alpha \lambda_j^\beta$, see also \eqn{xthetadef}. The denominator factors in eq.~(\ref{x2p}) drop out of any dual conformally invariant quantity and so may be ignored for our purposes.
We will use eq.~(\ref{x2p}) to establish the connection between the cluster $\mathcal{A}$-coordinates (defined in terms of Pl\"ucker coordinates) and the Steinmann relations (formulated in terms of Mandelstam invariants). More general Pl\"ucker coordinates $\ket{ijkl}$ not of the form $\ket{i{-}1\,i\,j{-}1\,j}$ have more complicated (algebraic) representations in terms of Mandelstam invariants. (A systematic approach for finding such representations was discussed in the appendix of ref.~\cite{Prygarin:2011gd}.)

In this paper we focus on $n=7$ where there are a finite number of $\mathcal{A}$-coordinates. In addition to the Pl\"ucker coordinates $\ket{ijkl}$ there are 14 Pl\"ucker bilinears of the form $\ket{a(bc)(de)(fg)} \equiv \ket{abde}\ket{acfg}-\ket{abfg}\ket{acde}$.
A convenient complete and multiplicatively independent set of 42 dual conformally invariant ratios, introduced in ref.~\cite{Drummond:2014ffa}, is given in terms of these building blocks by
\begin{align}
\label{Salphabet}
a_{11} &= \frac{\ket{1234}\ket{1567}\ket{2367}}{\ket{1237}\ket{1267}\ket{3456}}\,, \quad & \quad
a_{41} &= \frac{\ket{2457}\ket{3456}}{\ket{2345}\ket{4567}}\,, \nonumber \\
a_{21} &= \frac{\ket{1234}\ket{2567}}{\ket{1267}\ket{2345}}\,, \quad & \quad
a_{51} &= \frac{\ket{1(23)(45)(67)}}{\ket{1234}\ket{1567}}\,, \\
a_{31} &= \frac{\ket{1567}\ket{2347}}{\ket{1237}\ket{4567}}\,, \quad & \quad
a_{61} &= \frac{\ket{1(34)(56)(72)}}{\ket{1234}\ket{1567}}\,, \nonumber
\end{align}
with $a_{ij}$ for $1< j \le 7$ given by cyclic permutation of the particle labels; specifically,
\begin{equation}
\label{Salphabetpermutation}
a_{ij} = a_{i1}\big |_{Z_{k} \rightarrow Z_{k+j-1}}\,.
\end{equation}
The Steinmann relations, to be reviewed in section~\ref{Section:SteinmannRelations}, are expressed simply in terms of Mandelstam invariants. We therefore note that with the help of eq.~(\ref{x2p}) we can express $a_{1j}$ quite simply as
\begin{equation}\label{a11tos}
a_{11} = \frac{s_{23}s_{67}s_{712}}{s_{12}s_{71}s_{45}}\,,
\end{equation}
with the remaining six $a_{1j}$ again given by cyclic permutations. The remaining 35 cluster $\mathcal{A}$-coordinates do not admit simple representations in terms of Mandelstam invariants because they involve brackets not of the form $\ket{i{-}1\,i\,j{-}1\,j}$.

Finally, it is useful to relate the cross ratios $u_i$, defined in \eqn{uidef}, to the letters $a_{ij}$.  Eq.~\eqref{a11tos} can alternatively be written as
\begin{equation}\label{a11tox}
a_{11} = \frac{x_{24}^2x_{61}^2x_{73}^2}{x_{13}^2x_{72}^2x_{46}^2}\,.
\end{equation}
Combining this equation with cyclic permutations of it, and using \eqn{uidef}, we find that
\begin{equation}\label{a11a14a15toxtou}
\frac{a_{11}}{a_{14}a_{15}} = \frac{x_{73}^2x_{46}^2}{x_{74}^2x_{36}^2} = u_{36} = u_2 \,,
\end{equation}
plus cyclic permutations of this relation. Note that, although we can define 7 of these cross ratios $u_i$ in seven-point kinematics, an $n$-point scattering process in this theory only has $3n-15$ algebraically independent dual conformal invariants.  Thus only 6 of the 7 $u_i$ (or $a_{1i}$) are algebraically independent.  The seven $u_i$ obey a single algebraic equation, the condition that a particular Gram determinant vanishes, which restricts the kinematics to a six-dimensional surface within the seven-dimensional space of cross ratios. We will not need the explicit form of the Gram determinant in this paper.

\subsection{Integrability}
\label{Section:Integrability}

The heptagon bootstrap is based on the working hypothesis that any seven-point $L$-loop amplitude in planar $\mathcal{N}=4$ SYM theory can be expressed as a linear combination of weight-$2L$ generalized polylogarithm functions written in the 42-letter alphabet shown in eq.~(\ref{Salphabet}). Using this alphabet one can write $42^{k}$ distinct symbols of weight $k$. Fortunately, relatively few linear combinations of these $42^{k}$ symbols are actually the symbol of some function. A symbol $\mathcal{S}$ of the form
\begin{equation}
\mathcal{S}(f_{k}) = \sum_{\alpha_{1},\ldots,\alpha_{k}} f_{0}^{(\alpha_{1},\ldots,\alpha_{k})} (\phi_{\alpha_{1}} \otimes \cdots \otimes \phi_{\alpha_{k}}),
\end{equation}
where the $\phi_{\alpha_{j}}$ are letters, corresponds to an actual function only if it satisfies the integrability condition
\begin{equation}
\label{eq:integrability}
\sum_{\alpha_{1},\ldots,\alpha_{k}} f_{0}^{(\alpha_{1},\ldots,\alpha_{k})} \underbrace{(\phi_{\alpha_{1}} \otimes \cdots \otimes \phi_{\alpha_{k}})}_{\text{omitting } \alpha_{j} \otimes\alpha_{j{+}1}} d\text{log} \phi_{\alpha_{j}} \wedge d\text{log} \phi_{\alpha_{j{+}1}} = 0 \quad \forall j \in \{1,2,\ldots, k{-}1\}\,.
\end{equation}
A conceptually simple method for determining all integrable symbols of a given weight $k$ is discussed in appendix~\ref{Appendix:BootstrapApproach}, where the definition of the wedge product appearing in the above equation is also given. The symbols of physical amplitudes have several additional properties, to which we will now turn our attention.

\subsection{Symbol Singularity Structure}
\label{Section:SymbolSingularityStructure}

Locality requires that amplitudes can only have singularities when an intermediate particle goes on-shell. In a planar theory the momenta of intermediate particles can always be expressed as a sum of cyclically adjacent momenta, and thresholds in massless theories are always at the origin.  Hence perturbative amplitudes in planar SYM theory can only have branch points when the corresponding Mandelstam invariants $s_{i,\ldots,j-1} = x_{ij}^2$ vanish.

When some letter $\phi$ appears in the first entry of a symbol it indicates that the corresponding function has branch points at $\phi = 0$ and $\phi = \infty$. Therefore the first entry of a symbol that corresponds to a physical scattering amplitude must be a ratio of products of $x_{ij}^2$~\cite{Gaiotto:2011dt}. We see from eqs.~\eqref{x2p} and~\eqref{Salphabet} that only the seven $a_{1j}$ are valid first entries. The remaining 35 cluster $\mathcal{A}$-coordinates contain terms that may be zero (or infinite) without any intermediate particles going on-shell. There is no possibility of cancellation in a sum over terms in a symbol since the letters of the alphabet are multiplicatively independent. The restriction that the first entry of the symbol of any seven-point amplitude must be one of the seven $a_{1j}$ is called the first-entry condition.

\subsection{Steinmann Relations}
\label{Section:SteinmannRelations}

\begin{figure}
\center
\includegraphics[width=0.75\linewidth]{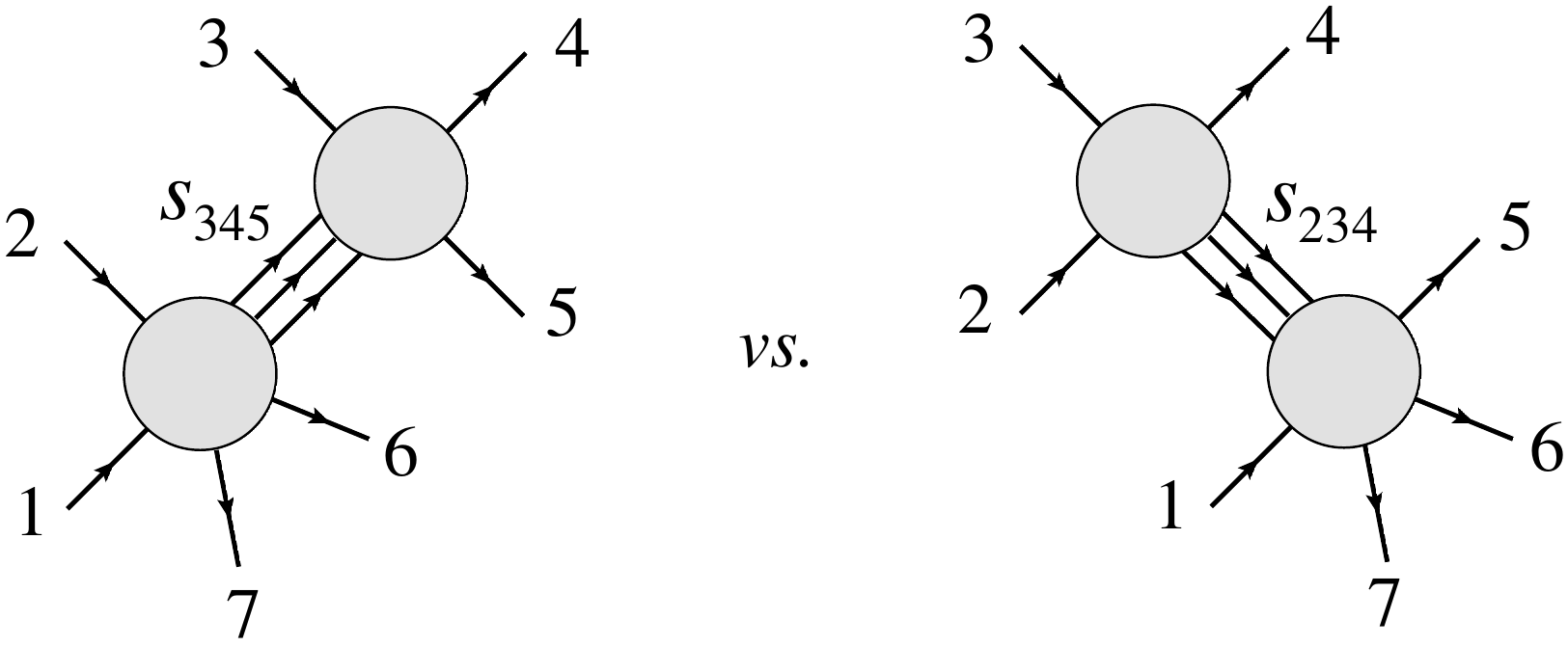}
\caption{The figure on the left (right) shows the discontinuity of an amplitude in the $s_{345}$ ($s_{234}$) channel due to the respective intermediate states.  These two channels overlap, which implies that the states that cross the first cut cannot produce a discontinuity in the second channel (or vice versa).}
\label{Fig:Steinmann}
\end{figure}

The classic work of Steinmann provided powerful restrictions on the analytic form of discontinuities~\cite{Steinmann}. Expanding upon his work, Cahill and Stapp found that the generalized Steinmann relations hold and that double discontinuities vanish for any pair of overlapping channels~\cite{Cahill:1973qp}.\footnote{The implications of the Steinmann relations for the multi-Regge limit of amplitudes in planar $\mathcal{N}=4$ SYM have been analyzed in refs.~\cite{Brower:2008ia,Brower:2008nm,Bartels:2008ce,Bartels:2008sc}.} A channel is labelled by a Mandelstam invariant, but it also corresponds to an assignment of particles to incoming and outgoing states.  Two channels overlap if the four sets into which they divide the particles -- (incoming,incoming), (incoming,outgoing), (outgoing,incoming) and (outgoing,outgoing) -- are all non-empty.  \Fig{Fig:Steinmann} shows a pair of overlapping channels for the seven-point process, $s_{345}$ and $s_{234}$.  They overlap because they divide the seven particles into the four non-empty sets $\{2\}$, $\{3,4\}$, $\{5\}$, and $\{6,7,1\}$.

Unlike two-particle invariants, three-particle invariants can cross zero ``gently'', without any other invariants having to change sign.  \Fig{Fig:Steinmann} is drawn for the $3\to4$ configuration with particles 1, 2 and 3 incoming.  Within that configuration, the left panel shows that $s_{345}$ can be either negative or positive.  As $s_{345}$ moves from negative to positive, a branch cut opens up, due to one or more on-shell particles being allowed to propagate between the two blobs.  The discontinuity in the amplitude across the branch cut is given by the sum of all such on-shell intermediate-state contributions, integrated over their respective phase space.  The same is true for the $s_{234}$ discontinuity illustrated in the right panel.  However, once one takes the $s_{345}$ discontinuity, the resulting function cannot have a second discontinuity in the $s_{234}$ channel, because it is impossible for states to propagate on-shell simultaneously in both the $s_{345}$ and $s_{234}$ ``directions''.  Thus we require the Steinmann conditions,
\be
\label{eq:Steinmann7s}
\text{Disc}_{s_{i+1,i+2,i+3}}\left[\text{Disc}_{s_{i,i+1,i+2}}F \right]
= \text{Disc}_{s_{i+2,i+3,i+4}}\left[\text{Disc}_{s_{i,i+1,i+2}}F \right] = 0,
\ee
to hold for all $i=1,2,\ldots7$.

In contrast, the $s_{234}$ channel does not overlap the $s_{567}$ channel (or the $s_{671}$ channel).  For example, in the right panel of the figure, one can have a second discontinuity, after taking $\text{Disc}_{s_{234}}$, in the $s_{567}$ channel, as particle 1 and the particles crossing the $s_{234}$ cut rescatter into another set of intermediate states, which then materializes into particles 5, 6 and 7.  That is, the following double discontinuities can be nonvanishing,
\be\label{eq:Steinmann7not}
\text{Disc}_{s_{i+3,i+4,i+5}}\left[\text{Disc}_{s_{i,i+1,i+2}}F \right] \neq 0, \qquad
\text{Disc}_{s_{i+4,i+5,i+6}}\left[\text{Disc}_{s_{i,i+1,i+2}}F \right] \neq 0,
\ee
and they provide us with no useful constraints.  Also, the ``self'' double discontinuities are nonvanishing,
\be\label{eq:Steinmann7selfnot}
\text{Disc}_{s_{i,i+i,i+2}}\left[\text{Disc}_{s_{i,i+1,i+2}}F \right] \neq 0,
\ee
and are not of use to us.  A recent analysis of the Steinmann relations, focusing on the six-point case, can be found in ref.~\cite{Caron-Huot:2016owq}.

We will only consider restrictions imposed on the symbol letters $a_{ij}$ by the Steinmann relations on overlapping three-particle cuts, \eqn{eq:Steinmann7s}. If there are any restrictions imposed by using two-particle cuts, they are considerably more subtle for generic kinematics. Flipping the sign of a two-particle invariant generally entails moving a particle from the initial state to the final state, or vice versa, and other invariants can flip sign at the same time, making it hard to assess the independence of the two-particle discontinuities.

Because the discontinuities of a symbol are encoded in its first entries, double discontinuities are encoded by the combinations of first and second entries that appear together. Correspondingly, the Steinmann relations tell us that the symbol of an amplitude cannot have any terms in which overlapping three-particle Mandelstam invariants appear together as first and second entries. Eqs.~\eqref{x2p}--\eqref{Salphabet} imply that this only imposes a constraint on the letters $a_{1j}$, since the other letters do not contain three-particle Mandelstam invariants $s_{i{-}1,i,i+1}\propto \langle i{-}2\,i{-}1\,i{+1}\,i{+}2\rangle$. More specifically, we see in~\eqn{a11tos} that each $a_{1i}$ is proportional to a single three-particle invariant $s_{i{-}1,i,i+1}$, so a first entry of $a_{1i}$ cannot be followed by a second entry of $a_{1,i+1}, a_{1,i+2}, a_{1,i+5}$, or $a_{1,i+6}$, all of which contain a three-particle invariant involving $p_{i{-}1}$, $p_{i}$, or $p_{i+1}$.  A first entry of $a_{1i}$ {\it can} be followed by a second entry of $a_{1i}$, $a_{1,i+3}$, $a_{1,i+4}$, or any $a_{ki}$ for $k>1$ (subject to the constraint of integrability).

Everything stated thus far about the Steinmann constraint applies to full, infrared-divergent amplitudes. However, the BDS-like-normalized amplitudes straightforwardly inherit this constraint, due to the fact that the BDS-like ansatz, given explicitly in eqs.~(\ref{hatM7}) and (\ref{A7BDSlike}), contains no three-particle invariants; it therefore acts as a spectator when taking three-particle discontinuities, e.g.
\be
\text{Disc}_{s_{i{-}1,i,i+1}} \cA^{\text{MHV}}_7
= \text{Disc}_{s_{i{-}1,i,i+1}} \Bigl[ \cA^{\text{BDS-like}}_7 \, \cE \Bigr]
= \cA^{\text{BDS-like}}_7\ \text{Disc}_{s_{i{-}1,i,i+1}} \cE \,.
\ee
This is no longer true for the BDS-normalized amplitude, which according to \eqn{eq:BDS-like_def} comes with an extra factor of $\exp[\frac{\Gamma_{\text{cusp}}}{4} Y_n]$. When expanded at weak coupling this factor will produce powers of $Y_n$. The function $Y_n$ is itself Steinmann since $Y_n = - \cE^{(1)}_n$.  However, products of Steinmann functions are not generically Steinmann functions, because overlapping discontinuities can arise from different factors in the product.  Indeed, once we observe that $Y_n$ has a cut in one three-particle channel, and that it is dihedrally invariant, we know it has cuts in all three-particle channels. Whereas $Y_n$ itself is a \emph{sum} of terms having cuts in overlapping channels, it is the \emph{cross terms} in $(Y_n)^2$, or higher powers of $Y_n$, that violate the Steinmann relations.   Similarly, the ratio function $V_* = E_*/\cE$, when expanded out perturbatively, contains products of Steinmann functions and therefore does not obey the Steinmann relations.  The lesson here is that the proper normalization of the amplitude is critical for elucidating its analytic properties.

To summarize, the Steinmann relations require that any BDS-like-normalized seven-point function $F$, such as $\cE_7$ or $E_7$, must satisfy
\be\label{eq:Steinmann7}
\text{Disc}_{a_{1i}}\left[\text{Disc}_{a_{1j}}F \right] = 0
\quad \text{if }\,j\ne i,i+3,i+4\,.
\ee
At the level of the symbol, this statement is equivalent to requiring that the symbol of $F$ contains no first entries $a_{1i}$ followed by second entries $a_{1,i+1}$, $a_{1,i+2}$, $a_{1,i+5}$, or $a_{1,i+6}$.

\subsection{Absence of Triple Discontinuity Constraints}
\label{Section:NoTripleDiscConstraints}

At the seven-point level, it is interesting to ask whether there could be new constraints on amplitudes of the following type:
\be\label{eq:Triple}
\text{Disc}_{a_{17}}\biggl[
\text{Disc}_{a_{14}}\Bigl[\text{Disc}_{a_{11}}F \Bigr] \biggr] \ \stackrel{?}{=}\ 0.
\ee
The three-particle channels corresponding to $a_{11}$ and $a_{14}$ do not overlap,
nor do the channels corresponding to $a_{14}$ and $a_{17}$.
The channels corresponding to $a_{11}$ and $a_{17}$ \emph{do} overlap, but
the two discontinuities are separated by the $a_{14}$ discontinuity in
between.  (An analogous situation never arises for three-particle cuts
in the six-point case, because the only allowed double three-particle cut
in that case involves cutting the same invariant twice.)
We have inspected the symbols of the MHV and NMHV seven-point amplitudes, and we find that \eqn{eq:Triple} is generically non-vanishing.  The act of taking the non-overlapping second discontinuity of the amplitude apparently alters the function's properties enough that the third discontinuity is permitted.

\subsection{Steinmann Heptagon Functions}
\label{Section:SteinmannHeptagonFunctions}

We define a heptagon function of weight $k$ to be a generalized polylogarithm function of weight $k$ whose symbol may be written in the alphabet of 42 cluster $\mathcal{A}$-coordinates, eq.~(\ref{Salphabet}), and which satisfies the first entry condition. These functions have been studied in ref.~\cite{Drummond:2014ffa}, where it was found that the vector space of heptagon function symbols at weight $k=$ 1, 2, 3, 4, 5 has dimension 7, 42, 237, 1288, 6763, respectively.

In this paper our goal is to sharpen the heptagon bootstrap of ref.~\cite{Drummond:2014ffa} by taking advantage of the powerful constraint provided by the Steinmann relations. We thus define Steinmann heptagon functions to be those heptagon functions that additionally satisfy the Steinmann relations~\eqref{eq:Steinmann7}. This corresponds to a restriction on the second entry of their symbols, as discussed in section~\ref{Section:SteinmannRelations}. We stress again that while both BDS-normalized and BDS-like-normalized amplitudes are heptagon functions, only the BDS-like-normalized ones, $\cE$, $E_0$, and $E_{ij}$, are Steinmann heptagon functions.

We will see in subsection~\ref{Section:SteinmannHeptagonSymbolsandTheirProperties} that a drastically reduced number of heptagon functions satisfy the Steinmann relations.  The reduction begins at weight 2, where there are 42 heptagon function symbols, but only 28 that obey the Steinmann relations.
The corresponding 28 functions fall into 4 orbits:
\be
{\rm Li}_2\left(1-\frac{a_{13}a_{14}}{a_{17}}\right) \,, \quad
{\rm Li}_2\left(1-a_{14}a_{16}\right) \,, \quad
\log^2 a_{13} \,, \quad
\log a_{13}
\log a_{16} \,,
\label{SteinmannheptWt2}
\ee
together with their cyclic permutations. This fractional reduction, by one third, is the same as in the hexagon case~\cite{Caron-Huot:2016owq}, where the number of weight-2 functions was reduced from 9 to 6.  At higher weight, we will see that the reductions are much more dramatic, and even more so for heptagon functions than hexagon functions.  This reduction in the number of relevant functions vastly decreases the size of our ansatz, making this version of the bootstrap program more computationally tractable than its predecessor.


\section{MHV and NMHV Constraints}
\label{Section:MHVandNMHVConstraints}

In appendix~\ref{Appendix:BootstrapApproach} we provide an algorithm for generating a basis for the symbols of weight-$k$ Steinmann heptagon functions, which serve as ans\"atze for the MHV and NMHV amplitudes. We then impose known analytic and physical properties as constraints in order to identify the amplitudes uniquely. Here we review these properties and the constraints they impose.

\subsection{Final Entry Condition}
\label{Section:FinalEntryCondition}

The final entry condition is a restriction on the possible letters that may appear in the final entry of the symbol of an amplitude. As a consequence of the dual superconformal symmetry of SYM, the differential of an MHV amplitude must be expressible as a linear combination of $d\log\ket{i\, j{-}1\, j\, j{+}1}$ factors~\cite{CaronHuot:2011ky}. The differential of a generalized polylogarithm of weight $k$ factors into linear combinations of weight-$(k{-}1)$ polylogarithms multiplied by $d\log\phi$ terms where $\phi$ is the final entry of the symbol. Therefore the final entries of the symbol of an MHV amplitude must be composed entirely of Pl\"ucker coordinates with three adjacent momentum twistors, $\ket{i\, j{-}1\, j\, j{+}1}$. In the symbol alphabet~(\ref{Salphabet}) we have chosen, the final entries can only be drawn from the set of 14 letters $\{a_{2j},a_{3j}\}$.

The MHV final entry condition we just described can be derived from an anomaly equation for the $\bar{Q}$ dual superconformal generators~\cite{CaronHuot:2011kk}. The same anomaly equation can also be used to constrain the final entries of the symbol of the NMHV superamplitude $E$. In particular, using as input the leading singularities of the N$^2$MHV 8-point amplitude obtained from the Grassmannian~\cite{ArkaniHamed:2009dn}, and refining the $\bar{Q}$ equation so as to act on the BDS-like normalized amplitude rather than the BDS-normalized one, Caron-Huot has found~\cite{SCprivate} that only 147 distinct ($R$-invariant) $\times$ (final entry) combinations are allowed in $E$, namely these 21:
\begin{gather}\label{FinalEntryNMHV}
\text{(34)} \log a_{21}, \quad\text{(14)} \log a_{21}, \quad\text{(15)} \log a_{21}, \quad\text{(16)} \log a_{21}, \quad\text{(13)} \log a_{21}, \quad\text{(12)} \log a_{21},\nonumber\\
\quad\text{(45)} \log a_{37} ,\quad\text{(47)} \log a_{37},\quad\text{(37)} \log a_{37},\quad\text{(27)} \log a_{37} ,\quad\text{(57)} \log a_{37} ,\quad\text{(67)} \log a_{37},\,\,\,\,\,\nonumber\\
\text{(45)} \log \frac{a_{34}}{a_{11}}, \quad\text{(14)} \log \frac{a_{34}}{a_{11}}, \quad\text{(14)} \log \frac{a_{11} a_{24}}{a_{46}}, \quad\text{(14)} \log \frac{a_{14} a_{31}}{a_{34}},\\
\text{(24)} \log \frac{a_{44}}{a_{42}},\quad\quad \text{(56)} \log a_{57},\quad\quad \text{(12)} \log a_{57}, \quad\quad\text{(16)} \log \frac{a_{67}}{a_{26}},\nonumber\\
\text{(13)} \log \frac{a_{41}}{a_{26} a_{33}} +(\text{(14)}-\text{(15)}) \log a_{26} -\text{(17)} \log a_{26} a_{37} +\text{(45)} \log \frac{a_{22}}{a_{34} a_{35}} -\text{(34)} \log a_{33} \,,\nonumber
\end{gather}
together with their cyclic permutations.\footnote{We thank Simon Caron-Huot for sharing these results with us.}

\subsection{Discrete Symmetries}
\label{Section:DiscreteSymmetries}
The $n$-particle superamplitudes $\mathcal{A}_n$ are invariant under dihedral transformations acting on the external particle labels. The generators of the dihedral group $D_n$ are the cyclic permutation $i\to i+1$ and the flip permutation $i\to n+1-i$ of the particle labels, or equivalently of the momentum twistors. For the heptagon $a$-letters~\eqref{Salphabet}, these correspond to
\be
\begin{aligned}\label{eq:dihedrala}
\text{Cyclic transformation:}& \quad a_{li}\to a_{l,i+1}\,,\\
\text{Flip transformation:}& \begin{cases}
a_{2i}\leftrightarrow a_{3,8-i}\\
a_{li}\to a_{l,8-i}&\text{for}\quad l\ne 2,3\,.
\end{cases}
\end{aligned}
\ee

MHV and $\overline{\text{MHV}}$ amplitudes differ only in their tree-level prefactors. Hence the functions $\mathcal{E}_n$ and $R_n$ must remain invariant under spacetime parity transformations. Parity maps NMHV amplitudes to $\overline{\text{NMHV}}$ ones and therefore acts nontrivially on $E_0$, $E_{12}$ and $E_{14}$. In the language of our symbol alphabet (\ref{Salphabet}), a parity transformation leaves the letters $a_{1i}$ and $a_{6i}$ invariant. The remaining letters transform under parity according to
\begin{equation}
  \text{Parity transformation:}\quad a_{21} \longleftrightarrow a_{37},
                              \qquad a_{41} \longleftrightarrow a_{51},
\end{equation}
and the cyclic permutations thereof.

The parity and dihedral symmetries of the (super)amplitude are inherited by its BDS(-like) normalized counterpart because the BDS(-like) ans\"atze are also dihedrally invariant.

\subsection{Collinear Limit}
\label{Section:CollinearLimit}
So far we have primarily focused on the BDS-like normalized amplitude and the Steinmann functions describing it. However for the study of collinear limits it proves advantageous to switch, using \eqn{eq:BDSLikeToBDS}, to the BDS-normalized amplitude, since in the limit the former becomes divergent, whereas the latter remains finite.

In more detail, the BDS ansatz $A_n^{\text{BDS}}$ entering eq.~(\ref{eq:remainderdefinition}) is defined in such a way that the $n$-point BDS-normalized amplitude (or equivalently the remainder function for MHV) reduces to the same quantity but with one fewer particle:
\begin{equation}
\label{eq:collinear}
\begin{aligned}
\underset{i+1 || i}{\text{lim}} R_{n} = R_{n{-}1}\,,\\
\lim_{i{+}1 \parallel i} B_n = B_{n-1}\,.
\end{aligned}
\end{equation}
To take one of these collinear limits, one of the $s_{i,i+1}$ must be taken to zero. From eq.~(\ref{x2p}), we see that this can be accomplished by taking a limit of one of the momentum twistor variables. In the case of the NMHV superamplitude we also need to specify the limit of the fermionic part of the supertwistors~\eqref{eq:supertwistors}. The (MHV degree preserving) $7||6$ collinear limit can be taken by sending
\begin{equation}
\label{eq:Zcollinear}
\cZ_7 \to \cZ_6 + \epsilon \frac{\vev{1246}}{\vev{1245}} \cZ_5 +
\epsilon \tau \frac{\vev{2456}}{\vev{1245}} \cZ_1 +
\eta \frac{\vev{1456}}{\vev{1245}} \cZ_2\,,
\end{equation}
for fixed $\tau$, and by taking the limit $\eta \rightarrow 0$ followed by $\epsilon \rightarrow 0$. 

Of course for bosonic quantities, only the bosonic part $\cZ_i\to Z_i$ of the supertwistor is relevant. As noted in ref.~\cite{Drummond:2014ffa}, in the limit~\eqref{eq:Zcollinear} the heptagon alphabet~\eqref{Salphabet} reduces to the hexagon alphabet, plus the following 9 additional letters,
\begin{gather}
\eta\,,\quad \epsilon\,,\quad \tau\,,\quad 1 + \tau\,,
\cr
\vev{1235} \vev{1246} + \tau \vev{1236} \vev{1245}\,,\quad\vev{1245} \vev{3456} + \tau \vev{1345} \vev{2456}\,,
\cr
\vev{1246} \vev{2356} + \tau \vev{1236} \vev{2456}\,,\quad \vev{1246} \vev{3456} + \tau \vev{1346} \vev{2456}\,,
\cr
\vev{1235} \vev{1246} \vev{3456} + \tau \vev{1236} \vev{1345} \vev{2456}\,.
\label{eq:badletters}
\end{gather}
Therefore the collinear limits of heptagon functions are not generically hexagon functions. We say that a heptagon symbol has a well-defined $7||6$ limit only if in this limit it is independent of all 9 of the additional letters \eqref{eq:badletters}.

We must also take the limit~\eqref{eq:Zcollinear} of the $R$-invariants. Since these invariants are antisymmetric under the exchange of any pair of twistor indices, the invariants that contain both indices $6$ and $7$ will vanish. All other invariants reduce to six-point $R$-invariants. Denoting the six-point invariants by
\be\label{Rinvariant6}
[12345]=(6)
\ee
and its cyclic permutations (under the six-point dihedral group), and solving the single identity of type~\eqref{sixZidentity} among them to eliminate $(6)$, we deduce that
\be
\begin{aligned}\label{eq:Bcollinear}
\lim_{7 \parallel 6} B=&(1) [\hat B_{17} + \hat B_{67} + \hat B_0]+(2) [\hat B_{26} - \hat B_{67}] + (3) [\hat B_{36} + \hat B_{37} + \hat B_{67} + \hat B_0]\\
&+(4) [ \hat B_{47} - \hat B_{67}] + (5) [\hat B_{56} + \hat B_{67} + \hat B_0]\,,
\end{aligned}
\ee
where the hats denote the collinear limit of the corresponding bosonic functions.

Finally, we should note that in this work we will be focusing on collinear limits of dihedrally invariant functions. Therefore it will be sufficient to consider the $7||6$ limit shown above, and the remaining $i{+}1 \parallel i$ collinear limits will be automatically satisfied as a consequence of dihedral symmetry.


\section{Results}
\label{Section:Results}

\subsection{Steinmann Heptagon Symbols and Their Properties}
\label{Section:SteinmannHeptagonSymbolsandTheirProperties}

As defined in section~\ref{Section:SteinmannHeptagonFunctions}, a Steinmann heptagon function of weight $k$ is a polylogarithm of weight $k$ that has a symbol satisfying the following properties:
\begin{enumerate}
\item[(i)] it can be expressed entirely in terms of the heptagon symbol alphabet of \eqn{Salphabet},
\item[(ii)] only the seven letters $a_{1i}$ appear in its first entry,
\item[(iii)] a first entry $a_{1i}$ is not followed by a second entry $a_{1j}$ with $j\in \{i+1,i+2,i+5,i+6\}$.
\end{enumerate}
We will frequently use the term `Steinmann heptagon symbol' to mean the symbol of a Steinmann heptagon function. We begin by investigating how the number of Steinmann heptagon symbols compares to the number of heptagon symbols reported in ref.~\cite{Drummond:2014ffa} through weight 5.

\renewcommand{\arraystretch}{1.25}
\begin{table}[!ht]
\begin{center}
\begin{tabular}{|l|>{\hfill}p{.83cm}|>{\hfill}p{.83cm}|>{\hfill}p{.83cm}|>{\hfill}p{.83cm}|>{\hfill}p{.83cm}|>{\hfill}p{.83cm}|>{\hfill}p{.83cm}|>{\hfill}p{.83cm}|}
\hline \hline
\multicolumn{1}{|c|}{$~$ \hfill Weight $k=$}
&\multicolumn{1}{c|}{$1$}
&\multicolumn{1}{c|}{$2$}
&\multicolumn{1}{c|}{$3$}
&\multicolumn{1}{c|}{$4$}
&\multicolumn{1}{c|}{$5$}
&\multicolumn{1}{c|}{$6$}
&\multicolumn{1}{c|}{$7$}
&\multicolumn{1}{c|}{$7''$}\\
\hline\hline
parity $+$, flip $+$ & 4 & 16 & 48 & 154 & 467 & 1413 & 4163 & 3026 \\
\hline
parity $+$, flip $-$ & 3 & 12 & 43 & 140 & 443 & 1359 & 4063 & 2946 \\
\hline
parity $-$, flip $+$ & 0 & 0 & 3 & 14 & 60 & 210 & 672 & 668 \\
\hline
parity $-$, flip $-$ & 0 & 0 & 3 & 14 & 60 & 210 & 672 & 669 \\
\hline\hline
Total & 7 & 28 & 97 & 322 & 1030 & 3192 & 9570 & 7309 \\
\hline
\hline
\end{tabular}
\end{center}
\caption{Number of Steinmann heptagon symbols at weights 1 through 7, and those satisfying the MHV next-to-final entry condition at weight 7.} \label{tab:t1}
\end{table}

Table~\ref{tab:t1} presents the number of Steinmann heptagon symbols through weight 7, computed using the bootstrapping procedure outlined in appendix~\ref{Appendix:BootstrapApproach}. The total number of Steinmann symbols through weight 5 can be compared to $7, 42, 237, 1288$, and $6763$ linearly independent heptagon symbols at weights 1 through 5, respectively~\cite{Drummond:2014ffa}. By weight 5, the size of the Steinmann heptagon space has already been reduced by a factor of six compared to the size of the standard heptagon space!  (The corresponding reduction factor for hexagon symbols at weight 5 is only about 3.5.)

The total number of Steinmann heptagon symbols at each weight was calculated without imposing spacetime parity or dihedral symmetries. The first four rows show the number of Steinmann heptagon symbols that have the specified eigenvalue under the $\mathbb{Z}_2 \times \mathbb{Z}_2$ generators of parity and the dihedral flip symmetry. There are many more parity even ($\text{parity }+$) Steinmann heptagon functions than parity odd.  At each weight there are approximately the same number of $\text{flip }+$ as $\text{flip }-$. Up through weight 7, there are an equal number of $\text{flip }+$ and $\text{flip }-$ parity odd functions.

Table~\ref{tab:t1} has two columns for weight 7. The column $7''$ counts the number of weight 7 symbols that satisfy an additional constraint we call the MHV next-to-final entry condition. Paired with the MHV final entry condition, which requires the final entry of the symbol to be $a_{2j}$ or $a_{3j}$, integrability imposes an additional constraint that prohibits the seven letters $a_{6i}$ from appearing in the next-to-final entry of any MHV symbol. Symbols satisfying this additional constraint are useful for bootstrapping the four-loop MHV heptagon, to be discussed in subsection~\ref{Section:The4LoopMHVHeptagon} below.

The fact that there are many more parity-even than parity-odd Steinmann heptagon functions is also true in the hexagon case~\cite{Caron-Huot:2016owq}.  In that case, it is possible to give a closed-form construction of an infinite series of parity-even ``$K$'' functions.  The $K$ functions apparently saturate the subspace of Steinmann hexagon functions having no parity-odd letters.  This series of functions can also be repurposed, with appropriate arguments, to describe some, but not all, of the Steinmann heptagon symbols having no parity-odd letters.

Before concluding this section, let us emphasize that we are here counting integrable symbols, not functions.  We expect each such symbol to be completable into a function.  However, there are other functions (with vanishing symbol) obtained by multiplying lower-weight functions by multiple zeta values.  When we impose physical constraints on the full function space, parameters associated with these additional functions will also have to be determined.  On the other hand, sometimes the function-level constraints are more powerful than the symbol-level constraints.  As first observed in the case of the 3-loop MHV hexagon~\cite{Dixon:2011pw,Dixon:2013eka}, the number of $n$-gon functions obeying additional constraints, such as well-defined collinear limits, may be smaller than the number of the corresponding symbols.  That is, completing a symbol to a function with proper branch cuts may require adding to it functions of lower weight that don't have a well-defined collinear limit, even if the symbol does. We leave the problem of upgrading our heptagon bootstrap from symbol to function level to a later work.

\subsection{The Three-Loop NMHV Heptagon}
\label{Section:The3LoopNMHVHeptagon}

Once we have constructed the Steinmann heptagon symbol space, we can assemble it into an ansatz for the seven-particle amplitude and apply the constraints outlined in section~\ref{Section:MHVandNMHVConstraints} to fix the free parameters. Let us describe the steps of this computation in the NMHV case.
\renewcommand{\arraystretch}{1.25}
\begin{table}[!ht]
\begin{center}
\begin{tabular}{|l|>{\hfill}p{1.46cm}|>{\hfill}p{1.46cm}|>{\hfill}p{1.46cm}|>{\hfill}p{1.46cm}|}
\hline \hline
\multicolumn{1}{|c|}{$~$ \hfill Loop order $L=$}
&\multicolumn{1}{c|}{$1$}
&\multicolumn{1}{c|}{$2$}
&\multicolumn{1}{c|}{$3$}\\
\hline\hline
Steinmann symbols & $15\times 28$ & 15$\times 322$ & $15\times 3192$\\
\hline
NMHV final entry& 42 & 85 & 226 \\
\hline
Dihedral symmetry& 5 & 11 & 31 \\
\hline
Well-defined collinear & 0 & 0 & 0 \\
\hline
\end{tabular}
\end{center}
\caption{Number of free parameters after applying each of the constraints in the leftmost column, to an ansatz for the symbol of the $L$-loop seven-point NMHV BDS-like-normalized amplitude. The first row in column $L$ is equal to the last line of column $k=2L$ of table~\ref{tab:t1}, multiplied by 15 for the 15 linearly independent $R$-invariants.} \label{tab:t2}
\end{table}

The NMHV amplitude is a linear combination of 15 transcendental functions multiplying the independent $R$-invariants.  Therefore the initial number of free parameters at $L$ loops, shown in table~\ref{tab:t2}, is given by 15 times the entry in table~\ref{tab:t1} that counts the total number of Steinmann heptagon symbols of weight $2L$.\footnote{If we had imposed dihedral symmetry first, we would have had only three independent functions $E_0$, $E_{12}$ and $E_{14}$ to parametrize, each with some dihedral symmetry, and there would have been fewer than 3 times the number of independent Steinmann heptagon symbols in the first line of the table. This part of the computation is not a bottleneck either way. This alternative procedure would also give rise to a different set of numbers in the second line of table~\ref{tab:t2}.}

We then impose the heptagon NMHV final entry condition discussed in subsection~\ref{Section:FinalEntryCondition}. Similarly to the NMHV hexagon case~\cite{Dixon:2015iva}, the list of allowed final entries in~\eqn{FinalEntryNMHV} can be translated into relations between the 42 different $\{k-1,1\}$ coproduct components for each of the 15 functions multiplying the independent $R$-invariants, for a total of 42$\times$15 = 630 independent objects. Note that \eqn{FinalEntryNMHV} contains all 21 distinct $R$-invariants, so in order to obtain the aforementioned equations we first need to eliminate the dependent $R$-invariants with the help of eqs.~\eqref{eq:P7tree} and~\eqref{eq:DependentRInvariant}.

In principle, one can impose the NMHV final entry equations at $L=k/2$ loops on the ansatz of weight-$k$ integrable symbols appearing in the first line of table~\ref{tab:t2}. In practice, we have found it more efficient to solve these equations simultaneously with the weight-$k$ integrability equations~\eqref{eq:integrability}, namely the equations imposing integrability on the last two slots of an ansatz for $E$. The number of free parameters after imposing this condition (using either method) is reported in the second line of table~\ref{tab:t2}. We see that the final entry condition is already very restrictive; out of the 47880 possible NMHV symbols with generic final entry at three loops, only 226 of them obey the NMHV final entry. Next we impose invariance of $E$ under dihedral transformations, as discussed in subsection~\ref{Section:DiscreteSymmetries}. The dihedral restriction leads to the small number of remaining free parameters reported in the third line of table~\ref{tab:t2}.

We then examine the behavior of the amplitude in the collinear limit. To this end, we recall from subsection~\ref{Section:CollinearLimit} that it is advantageous to convert to the BDS normalization, since the BDS-normalized amplitude is finite in the collinear limit, while the BDS-like normalized one becomes singular. Converting our partially-determined ansatz for $E$ to an equivalent ansatz for $B$ with the help of eq.~\eqref{eq:BDSLikeToBDS}, we then take its collinear limit using~\eqn{eq:Zcollinear}.

Quite remarkably, demanding that the right-hand side of~\eqn{eq:Bcollinear} be well-defined, namely independent of the spurious letters~\eqref{eq:badletters} (and thus also finite), suffices to uniquely fix $B$ through 3 loops!  Even an overall rescaling is not allowed in the last line of table~\ref{tab:t2}, because the condition of well-defined collinear limits, while homogeneous for BDS-normalized amplitudes, is inhomogeneous for the BDS-like normalization with which we work.  We did not need to require that the collinear limit~(\ref{eq:Bcollinear}) of the solution agrees with the six-point ratio function computed at three loops in ref.~\cite{Dixon:2014iba}, but of course we have checked that it does agree.

In this manner, we arrive at a unique answer for the symbol of the NMHV heptagon through three loops. Our results can be downloaded in a computer-readable file from~\cite{HeptagonsPage}. The one- and two-loop results match the amplitudes computed in refs.~\cite{Drummond:2008bq} and~\cite{CaronHuot:2011kk}, respectively. The fact that six-point boundary data is not even needed to fix the symbol through three loops points to a strong tension between the Steinmann relations, dual superconformal symmetry (in the guise of the final entry condition), and the collinear limit.

\subsection{The Four-Loop MHV Heptagon}
\label{Section:The4LoopMHVHeptagon}

For the MHV remainder function at $L=k/2$ loops, we could in principle start from an ansatz for $\cE^{(L)}_7$ involving all heptagon Steinmann symbols of weight $k$. As with the NMHV case, however, it is simpler to impose the MHV final-entry condition discussed in section~\ref{Section:FinalEntryCondition} at the same time as integrability on the last two entries of the symbol. In fact, our initial four-loop MHV ansatz was constructed using not just the MHV final-entry condition, but also the MHV next-to-final entry condition discussed in section~\ref{Section:SteinmannHeptagonSymbolsandTheirProperties}.

\renewcommand{\arraystretch}{1.25}
\begin{table}[!ht]
\begin{center}
\begin{tabular}{|l|>{\hfill}p{1.46cm}|>{\hfill}p{1.46cm}|>{\hfill}p{1.46cm}|>{\hfill}p{1.46cm}|p{1.46cm}|>{\hfill}p{1.46cm}|}
\hline \hline
\multicolumn{1}{|c|}{$~$ \hfill Loop order $L=$}
&\multicolumn{1}{c|}{$1$}
&\multicolumn{1}{c|}{$2$}
&\multicolumn{1}{c|}{$3$}
&\multicolumn{1}{c|}{$4$}\\
\hline\hline
Steinmann symbols & 28 & 322 & 3192 & ? \\
\hline
MHV final entry & 1 & 1 & 2 & 4 \\
\hline
Well-defined collinear & 0 & 0 & 0 & 0 \\
\hline
\end{tabular}
\end{center}
\caption{Free parameter count after applying each of the constraints in the leftmost column to an ansatz for the symbol of the $L$-loop seven-point MHV BDS-like-normalized amplitude.} \label{tab:t3}
\end{table}

In the first line of table~\ref{tab:t3}, we reiterate the number of Steinmann heptagon functions with general final entry.  In the second line of the table, we report the number of symbols that satisfy the MHV final entry condition. Clearly, there are only a few Steinmann heptagon functions at each weight that satisfy even these few constraints.  Note that we have not even imposed dihedral invariance, nor that the symbol have even spacetime parity.

To determine the third line of the table, we convert the ansatz to one for the BDS normalized amplitude, using \eqn{eq:BDSLikeToBDS} and the symbol of $Y_7$.  We then ask that this quantity have a well-defined collinear limit.  As in the NMHV case, there is a unique solution to this constraint, this time through four loops, as reported in the last line of table~\ref{tab:t3}; this unique solution must be the symbol of $\cE_7^{(L)}$.  Our results can be downloaded in computer-readable files from~\cite{HeptagonsPage}. Again the overall normalization is fixed because the last constraint is an inhomogeneous one for a BDS-like normalized amplitude.  The symbols of the two- and three-loop seven-point BDS remainder functions $R_7^{(2)}$, $R_7^{(3)}$ are known~\cite{CaronHuot:2011ky,Drummond:2014ffa}. We have converted these quantities to the BDS-like normalization with the help of eq.~\eqref{eq:ERrelations}, and they agree with our unique solutions.  At four loops, when we convert our unique solution for $\mathcal{E}_{7}^{(4)}$ (which has 105,403,942 terms) to $R_{7}^{(4)}$ (which has 899,372,614 terms), we find that its well-defined collinear limit agrees perfectly with the symbol of the four-loop six-point MHV remainder function $R_6^{(4)}$ computed in ref.~\cite{Dixon:2014voa}. Because we did not need to impose dihedral invariance, nor spacetime parity, we can say that even less input is needed to fix the symbol of the MHV amplitude through four loops than was needed for the three-loop NMHV amplitude!

Before concluding, let us note that although we used the Steinmann constraint to tightly constrain the space of symbols through which we had to sift in order to find the four-loop MHV heptagon, it is possible that the same result could have been obtained (in principle, with much more computer power), without it. In the second row of table~\ref{tab:t3} we see, for example, that at weight 6 there are precisely 2 Steinmann heptagon symbols satisfying the MHV final-entry condition. Ref.~\cite{Drummond:2014ffa} imposed the MHV final-entry condition, without considering the Steinmann relations, and found 4 different symbols at weight 6:  $(Y_7)^3$, $Y_7\,R_7^{(2)}$, $R_7^{(3)}$ and one more.  Modulo the reducible (product) functions $(Y_7)^3$ and $Y_7\,R_7^{(2)}$, heptagon functions satisfying the MHV final-entry condition automatically satisfy the Steinmann relations as well, at least at weight 6!  We cannot rule out the possibility that the Steinmann constraint is also superfluous at weight 8 (or, perhaps, even higher), but certainly the complexity of the computation is significantly reduced if one allows oneself to input this knowledge.

\subsection{Three Loops from Dihedral Symmetry}

In this subsection we consider dropping the final entry condition, which derives from dual superconformal invariance.  One motivation for doing this is to check independently the NMHV final entry conditions detailed in \eqn{FinalEntryNMHV}. Another possible motivation, in the MHV case, is to try to widen the applicability of the bootstrap approach to the study of (bosonic) light-like Wilson loops in weakly-coupled conformal theories with less supersymmetry than $\mathcal{N}=4$ SYM.

Let us consider adding general $L$-loop Steinmann heptagon symbols $\tilde{\cE}_7^{(L)}$ (with no restrictions on the final entry) to the known answer $\cE_7^{(L)}$ and see whether we can preserve the conditions of dihedral symmetry and good collinear behavior. We can ask this question through three loops, because we have a complete basis of Steinmann heptagon symbols up to (and beyond) weight six. Since such symbols appear additively in the BDS-normalized quantity $\cB_7^{(L)}$, we need the Steinmann symbols $\tilde{\cE}_7^{(L)}$ themselves to be well-defined in the collinear limit. The numbers of Steinmann heptagon symbols obeying the successive conditions of cyclic invariance, flip symmetry, and well-defined collinear behavior are detailed in table~\ref{tab:t4}. 

We find that the first dihedrally invariant Steinmann symbol with well-defined collinear limits appears at weight six, i.e.~at three loops. We denote this symbol by $\tilde{\cE}_7$.
In fact the collinear limit of $\tilde{\cE}_7$, which we denote by $\tilde{\cE}_6$, automatically turns out to possess six-point dihedral invariance as well. Furthermore the collinear limit of $\tilde{\cE}_6$ from six points to five is vanishing. Therefore the symbol $\tilde{\cE}_7$ could be added to that for $\cE_7^{(3)}$ (and simultaneously $\tilde{\cE}_6$ to $\cE_6^{(3)}$) without breaking dihedral symmetry or good collinear behavior either at seven points or at six points. 

Neither $\tilde{\cE}_7$ nor $\tilde{\cE}_6$ obey the MHV final entry condition, as required to be consistent with the results of section~\ref{Section:The4LoopMHVHeptagon}. Thus at the three-loop order, $\bar{Q}$-supersymmetry is really fixing only a single parameter, after the consequences of the Steinmann relations, dihedral symmetry and good collinear behavior are taken into account. A different criterion that can be used to uniquely determine $\cE_7^{(3)}$ is that the three-loop remainder $R_6^{(3)}$ should have at most a double discontinuity around the locus $u=0$ where $u$ is one of three the cross ratios available at six points. The double discontinuity is in fact predicted from the original implementation of the Wilson line OPE~\cite{Gaiotto:2011dt}, which we will not delve into here. We may simply observe that $\tilde{\cE}_6$ has a triple discontinuity and hence we can rule out adding $\tilde{\cE}_7$ to $\cE_7^{(3)}$ on these grounds.

\renewcommand{\arraystretch}{1.25}
\begin{table}[!ht]
\begin{center}
\begin{tabular}{|l|>{\hfill}p{1.46cm}|>{\hfill}p{1.46cm}|>{\hfill}p{1.46cm}|>{\hfill}p{1.46cm}|p{1.46cm}|>{\hfill}p{1.46cm}|}
\hline \hline
\multicolumn{1}{|c|}{$~$ \hfill Loop order $L=$}
&\multicolumn{1}{c|}{$1$}
&\multicolumn{1}{c|}{$2$}
&\multicolumn{1}{c|}{$3$}\\
\hline\hline
Steinmann symbols & 28 & 322 & 3192\\
\hline
Cyclic invariance & 4 & 46 & 456\\
\hline
Dihedral invariance & 4 & 30 & 255\\
\hline
Well-defined collinear & 0 & 0 & 1\\
\hline
\end{tabular}
\end{center}
\caption{Number of linearly independent Steinmann heptagon symbols obeying, respectively: cyclic invariance, dihedral invariance, and well-defined collinear behavior together with dihedral symmetry.} \label{tab:t4}
\end{table}

We may similarly examine the consequences of dihedral symmetry and collinear behavior for the NMHV amplitude. In this case there are some additional conditions which we can impose, from requiring the absence of spurious poles. We recall the form of the NMHV ratio function given in \eqn{eq:P7}, or equivalently the form of $E$ given in \eqn{eq:E7components}. The tree-level amplitude $\mathcal{P}^{(0)}$ obviously possesses only physical poles, but the individual $R$-invariants have spurious poles. Requiring that the NMHV amplitude as a whole has no spurious poles leads us to the following conditions:
\begin{align}
\text{Spurious I:} \quad  E_{47}|_{\langle 1356 \rangle =0} &= 0\,, \\
\text{Spurious II:} \quad  E_{23}|_{\langle 1467 \rangle =0} &= E_{25}|_{\langle 1467 \rangle =0}\,.
\end{align}

In table~\ref{tab:t5} we detail the number of Steinmann symbols obeying the successive conditions of cyclic symmetry, absence of spurious poles, well-defined collinear behavior, and flip symmetry. At weight two, we find a single combination obeying all conditions, which is precisely the combination $B^{(1)}$ itself, which is therefore determined up to an overall scale by these conditions. Note that unlike the $B^{(L)}$ for $L>1$, the function $B^{(1)}$ obeys the Steinmann relations.

At weight four, we find no Steinmann symbols obeying all the conditions. This is not in contradiction with the results of section~\ref{Section:The3LoopNMHVHeptagon}: we recall that the quantity $E^{(2)}$ does not exhibit well-defined, finite collinear behavior; rather it is the (non-Steinmann) function $B^{(2)}$ which manifests this. The zero in the final row of the $L=2$ column in table~\ref{tab:t5} rather reflects the fact that there is no Steinmann symbol which could be added to $E^{(2)}$ while preserving the good collinear behavior of $B^{(2)}$, even if we are willing to abandon the NMHV final entry condition.

At weight six, we find a single Steinmann symbol with all the properties listed in table~\ref{tab:t5}. It is precisely the same symbol $\tilde{\cE}_7$ appearing in table~\ref{tab:t4} multiplied by the tree-level amplitude $\mathcal{P}^{(0)}$.  Hence it only appears as a potential contribution to $E_0^{(3)}$. In other words, the symbols of $E_{12}^{(3)}$ and $E_{14}^{(3)}$ are uniquely fixed by the constraints of dihedral symmetry, absence of spurious poles and correct collinear behavior. The appearance of the same ambiguity $\tilde{\cE}_7$ in $E_0^{(3)}$ is to be expected since the only additional criterion imposed in table~\ref{tab:t5}, that of spurious-pole cancellation, cannot constrain potential contributions to $E_0$. Finally, we note that the addition of $\tilde{\cE}_7$ in $E_0^{(3)}$ is connected to its addition to $\cE_7^{(3)}$ by the NMHV to MHV collinear limit which relates $E_7$ to $\cE_6$. Thus dropping the final entry condition from $\bar{Q}$-supersymmetry allows only a single potential contribution at weight 6 in all of the heptagon and hexagon amplitudes.

\renewcommand{\arraystretch}{1.25}
\begin{table}[!ht]
\begin{center}
\begin{tabular}{|l|>{\hfill}p{2.46cm}|>{\hfill}p{2.46cm}|>{\hfill}p{3.46cm}|>{\hfill}p{2.46cm}|p{2.46cm}|>{\hfill}p{2.46cm}|}
\hline \hline
\multicolumn{1}{|c|}{$~$ \hfill Loop order $L=$}
&\multicolumn{1}{c|}{$1$}
&\multicolumn{1}{c|}{$2$}
&\multicolumn{1}{c|}{$3$}\\
\hline\hline
Steinmann symbols &$15\times 28$ & 15$\times 322$ & $15\times 3192$\\
\hline
Cyclic invariant& $4 + (2 \times 28)$ & $46 + (2 \times 322)$ & $456 + (2 \times 3192)$\\
\hline
Spurious vanishing I & 4 + 1 + 28 & 46 + 19 + 322 & 456 + 208 + 3192 \\
\hline
Spurious vanishing II & 4 + 6 & 46 + 89 & 456 + 927 \\
\hline
Well-defined collinear & 1 & 0 & 11\\
\hline
Flip invariant & 1 & 0 & 1\\
\hline
\end{tabular}
\end{center}
\caption{Number of Steinmann heptagon symbols entering the NMHV amplitude obeying respectively cyclic invariance, vanishing on spurious poles, well-defined collinear behavior and flip symmetry.} \label{tab:t5}
\end{table}

We conclude that, up to three loops, starting from an ansatz of Steinmann heptagon functions, all heptagon amplitudes and hence all hexagon amplitudes (by collinear limits) in planar $\mathcal{N}=4$ SYM can be determined just by imposing dihedral symmetry and well-defined collinear limits, combined with the requirement of no triple discontinuity in $R_6^{(3)}$ and no spurious poles in the NMHV amplitudes. These results provide an independent check of the NMHV final entry conditions~(\ref{FinalEntryNMHV}). It would be interesting to investigate whether the ambiguity functions $\tilde{\cE}_7$ and $\tilde{\cE}_6$ could play a role in the perturbative expansion of any weakly-coupled conformal theories with less supersymmetry than $\mathcal{N}=4$ SYM.


\section{The Multi-Particle Factorization Limit}
\label{Section:Multifact}

\begin{figure}
\center
\includegraphics[width=0.4\linewidth]{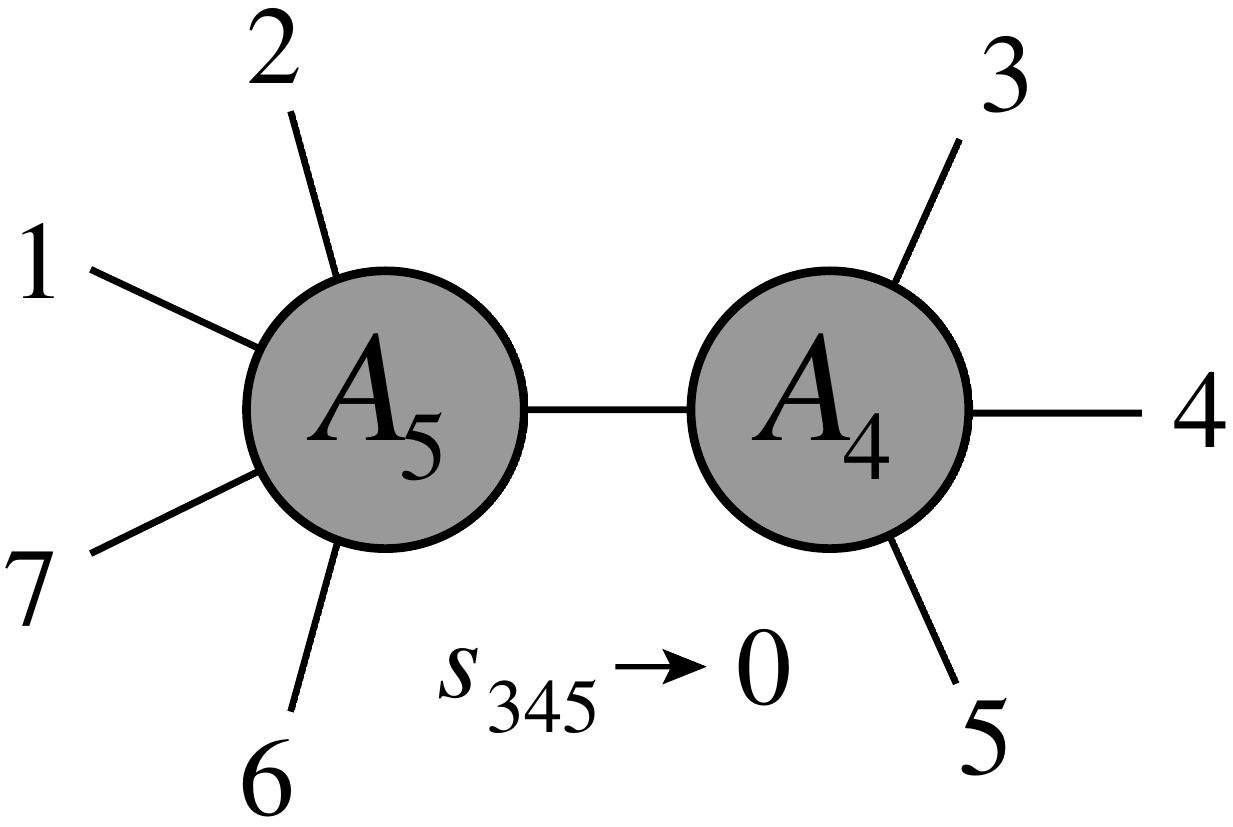}
\caption{Factorization of a seven-point amplitude in the limit $s_{345} {\rightarrow} 0$.  Notice that the collinear limit $p_7 \parallel p_1$ can be taken ``inside'' the factorization limit.}
\label{Fig:Multifact7}
\end{figure}

One of the kinematic limits we can study using our explicit seven-point results is the multi-particle factorization limit.  In this limit, one of the three-particle invariants goes on shell, $s_{i,i+1,i+2} \rightarrow 0$.  Figure~\ref{Fig:Multifact7} shows the limit $s_{345}\rightarrow0$.  In this limit the seven-point NMHV amplitude factorizes at leading power into a product of four-point and five-point amplitudes, multiplied by the $1/s_{345}$ pole.   The seven-point MHV amplitude vanishes at leading power.  Indeed, all supersymmetric MHV amplitudes are required to vanish at leading power when a three-particle (or higher-particle) invariant goes on shell.  This result holds because all possible helicity assignments for the intermediate state require at least one lower-point amplitude to have fewer than two negative-helicity gluons; such amplitudes vanish by supersymmetry Ward identities~\cite{Grisaru:1976vm,Grisaru:1977px}.  For the same reason, MHV tree amplitudes~\cite{Parke:1986gb} have no multi-particle poles.

Before turning to the behavior of the seven-point NMHV amplitude, we recall the multi-particle factorization behavior of the BDS-like-normalized six-point NMHV amplitude~\cite{Dixon:2014iba}.  As $s_{345}\rightarrow0$, two of the six-point $R$-invariants become much larger than the rest, and they become equal to each other.  Therefore the singular behavior of the six-point amplitude is controlled by a single coefficient function, which we denote by $\mathcal{U}_6$ and whose limiting behavior takes an especially simple form.\footnote{The function $\mathcal{U}_6$ can be identified with the function $E$ in refs.~\cite{Dixon:2015iva,Caron-Huot:2016owq}, but we prefer to adopt a different notation here to emphasize that this function is \emph{not} the BDS-like-normalized NMHV superamplitude $E_6$.} Up to power-suppressed terms, the limit of $\mathcal{U}_6$ was found to be a polynomial in $\log (uw/v)$, whose coefficients are rational linear combinations of zeta values, and whose overall weight is $2L$. Here, $u$, $v$, and $w$ are the three dual conformal invariant cross ratios for the hexagon, whose expressions in terms of six-point kinematics are
\be
u = \frac{x_{13}^2\,x_{46}^2}{x_{14}^2\,x_{36}^2}
  = \frac{s_{12}\,s_{45}}{s_{123}\,s_{345}} \,, \qquad
v = \frac{x_{24}^2\,x_{51}^2}{x_{25}^2\,x_{41}^2}
  = \frac{s_{23}\,s_{56}}{s_{234}\,s_{123}} \,, \qquad
w = \frac{x_{35}^2\,x_{62}^2}{x_{36}^2\,x_{52}^2}
   = \frac{s_{34}\,s_{61}}{s_{345}\,s_{234}} \,.  \label{uvw_def}
\ee
The six-point limit $s_{345} \rightarrow 0$ sends $uw/v \rightarrow \infty$.

The logarithm of $\mathcal{U}_6$, called $U$ in ref.~\cite{Dixon:2014iba}, has an even simpler behavior than $\mathcal{U}_6$. The $L$-loop contribution $U^{(L)}$ is also a polynomial in $\log (uw/v)$, but it has only degree $L$ at $L$ loops, for $L>1$.  This three-loop result was later found to hold also at four and five loops~\cite{Dixon:2015iva,Caron-Huot:2016owq}.  Because $U^{(L)}$ has weight $2L$, but a maximum of $L$ powers of $\log (uw/v)$ for $L>1$, every term in it contains zeta values, and its symbol vanishes.  The only exception is the one-loop result,
\be
U^{(1)}(u,v,w) \xrightarrow{\ s_{345}\rightarrow0 \ }
- \, \frac{1}{2} \log^2 \left(\frac{uw}{v}\right) - 2\zeta_2 \,,
\label{U6ptlimit}
\ee
where we have converted the result in ref.~\cite{Dixon:2014iba} to that for expansion parameter $g^2$.  The results for $U^{(L)}$ agree with the perturbative expansion of an all-orders prediction based on the Pentagon OPE~\cite{BSVprivate,SeverAmps15}.

Ref.~\cite{Dixon:2014iba} also made a prediction for the multi-particle factorization behavior of NMHV $n$-point amplitudes, which we can now test at 7 points at the symbol level.  Define the factorization function $F_n$ by
\begin{equation}
A_n^{\text{NMHV}}(k_i) \rightarrow
A_{j-i+1}(k_i, k_{i+1}, \dots, k_{j-1}, K) \frac{F_n(K^2,s_{l,l+1})}{K^2} A_{n-(j-i)+1}(-K, k_j, k_{j+1}, \dots, k_{i-1}) \,,
\end{equation}
as $K^2\to0$, or in the seven-point case,
\begin{equation}
A_7^{\text{NMHV}}(k_i) \xrightarrow{\ s_{345}\rightarrow0 \ }
A_5(k_6, k_7, k_1, k_2, K) \frac{F_7(K^2,s_{l,l+1})}{K^2} A_4(-K, k_3, k_4, k_5)\,,
\end{equation}
where $K=k_3+k_4+k_5$, $K^2=s_{345}$.
Then $F_7$ was predicted to have the form
\bea
[\log F_7]^{(L)}_{\rm symbol} &=& \delta_{L,1} \Biggl\{ \frac{1}{8\e^2}
  \biggl[ \biggl(\frac{(-s_{712})(-s_{34})}{(-s_{56})}\biggr)^{-\e}
   + \biggl(\frac{(-s_{45})(-s_{671})}{(-s_{23})}\biggr)^{-\e} \biggr]
\nonumber\\ &&\null\hskip1.0cm
  - \frac{1}{2} \log^2 \biggl(\frac{(-s_{712})(-s_{34})}{(-s_{56})}
        \bigg/ \frac{(-s_{45})(-s_{671})}{(-s_{23})}\biggr)
\nonumber\\ &&\null\hskip1.0cm
- \frac{1}{2} \log^2 \biggl(\frac{x_{73}^2 x_{35}^2 x_{46}^2 x_{62}^2}
                                 {x_{57}^2 x_{24}^2 (x_{36}^2)^2}\biggr)
\Biggr\} \,.
\label{lnF7Lsymb}
\eea
For simplicity, we have dropped all terms that vanish at symbol level, which
kills all terms in $\log F_7$ beyond one loop,
and we have converted to the $g^2$ expansion parameter.

We should now convert this prediction to one for the BDS-like normalized
amplitude.  Apart from trivial tree-level factors, we have
\be
\log F_7\ =\ 
\log\biggl(\frac{A_7^{\rm NMHV}}{A_5^{\rm BDS} \, A_4^{\rm BDS}}\biggr)
\ =\ \log\biggl(\frac{A_7^{\rm NMHV}}{A_7^{\rm BDS-like}}\biggr)
  - \log\biggl(\frac{A_5^{\rm BDS} \, A_4^{\rm BDS}}{A_7^{\rm BDS-like}}\biggr) \,.
\label{F7Ldef}
\ee
So to obtain $\log(A_7^{\rm NMHV}/A_7^{\rm BDS-like})$
we need to add to $[\log F_7]^{(1)}$ the quantity
\be
-\hat{M}_7^{(1)} + M_5^{(1)} + M_4^{(1)} \,,
\label{F7correction}
\ee
where $\hat{M}_7$ is given in \eqn{hatM7}, and $M_4^{(1)}$ and $M_5^{(1)}$ are the four- and five-point MHV amplitudes, for the kinematics shown in \fig{Fig:Multifact7}, and normalized by their respective tree amplitudes.

Adding eqs.~(\ref{F7Ldef}) and (\ref{F7correction}), we find, in terms of dual variables,
\be
\log\biggl(\frac{A_7^{\rm NMHV}}{A_7^{\rm BDS-like}}\biggr)^{(1)}
\! \to
- \frac{1}{2} \log^2 \biggl(\frac{x_{73}^2 x_{35}^2 x_{46}^2 x_{62}^2}
{x_{57}^2 x_{24}^2 (x_{36}^2)^2}\biggr)
- \frac{1}{2} \log^2\biggl(\frac{x_{46}^2 x_{72}^2 x_{13}^2}
                                  {x_{73}^2 x_{24}^2 x_{61}^2}\biggr)
- \frac{1}{2} \log^2\biggl(\frac{x_{35}^2 x_{72}^2 x_{61}^2}
                                  {x_{62}^2 x_{57}^2 x_{13}^2}\biggr) \,,
\label{lnA7Lsymb}
\ee
at symbol level, and a vanishing contribution to the logarithm beyond one loop.
Note that the first term in \eqn{lnA7Lsymb} comes directly out of \eqn{lnF7Lsymb}, and is the ``naive'' generalization of $-\frac{1}{2}\log^2(uw/v)$ to the seven-point case.  The first term diverges logarithmically as $s_{345} = x_{36}^2\to0$, while the last two terms are finite in this limit.

The one-loop factorization behavior in \eqn{lnA7Lsymb} could have been extracted, of course, from the one-loop seven-point amplitude.  Thus the symbol-level content of the prediction is really the vanishing of the logarithm beyond one loop.  Beyond symbol level, the all-loop-order prediction of ref.~\cite{Dixon:2014iba} is that (up to an additive constant) the first term gets upgraded to the function appearing in the six-point limit, namely $U(x)$, where $x = (x_{73}^2 x_{35}^2 x_{46}^2 x_{62}^2)/(x_{57}^2 x_{24}^2 (x_{36}^2)^2)$, while the last two terms should simply get multiplied by the cusp anomalous dimension.

Now let us test the symbol-level prediction~(\ref{lnA7Lsymb}) by taking the limit $s_{345} \rightarrow 0$ of the seven-point NMHV amplitude. Referring back to~\eqref{x2p}, we have
\begin{equation}
s_{345} = x_{36}^2 = \frac{\langle 23 56 \rangle}{\langle 23  \rangle \langle 56 \rangle} \rightarrow 0.
\end{equation}
Keeping $s_{23}$ and $s_{56}$ generic requires us to take this limit by sending $\langle 23 56 \rangle \rightarrow 0$. This limit can be accomplished using the replacement
\begin{align}
  \mathcal{Z}_2 \rightarrow \mathcal{Z}_3 + a \frac{\langle 1436 \rangle}{\langle 1456 \rangle } \mathcal{Z}_5 + b \frac{\langle 1453 \rangle}{\langle 1456 \rangle} \mathcal{Z}_6 + \epsilon \frac{\langle 3456 \rangle}{\langle 1456 \rangle } \mathcal{Z}_1
\label{Zfactlimit}  
\end{align}
where $a,b \in \mathbb{C}$ are generic and $\epsilon$ is a regulator. In the limit $\epsilon \rightarrow 0$, $a_{14}$ vanishes while the other $a_{ij}$ map into a space of 31 finite letters.

The map works out to be
\bea
a_{25} &\to& \frac{a_{11}a_{17}}{a_{21}a_{24}} \,, \qquad
a_{33} \to \frac{a_{17}}{a_{24}} \,, \qquad
a_{34} \to \frac{a_{21}a_{24}}{a_{17}} \,, \qquad
a_{37} \to \frac{a_{11}a_{17}}{a_{21}} \, \qquad
a_{42} \to a_{24} \,, \nonumber\\
a_{46} &\to& \frac{a_{21}a_{24}}{a_{17}} \,, \qquad
a_{52} \to \frac{a_{17}}{a_{24}} \,, \qquad
a_{56} \to \frac{a_{11}a_{17}}{a_{21}a_{24}} \,, \qquad
a_{63} \to -1, \qquad
a_{65} \to -1,
\label{afact}
\eea
which removes 10 of the 42 letters, leaving $a_{14}$ and the 31 finite letters.

We also need the limiting behavior of the seven-point $R$-invariants. Referring back to their definition~(\ref{fivebrak}), we see that the invariants (71), (14) and (47) become singular as $\langle 23 56 \rangle \rightarrow 0$ while all others remain finite. The finite $R$-invariants are suppressed in the identities \eqref{sixZidentity} in this limit, giving us
\begin{align}
(71)_{s_{345} \rightarrow 0}\ =\ (14)_{s_{345} \rightarrow 0}\ =\ (47)_{s_{345} \rightarrow 0} \,.
\label{factRbig}
\end{align}
The function controlling the behavior of $E_7$ as $s_{345} \rightarrow 0$ is thus given by the sum of functions multiplying these singular invariants in \eqn{eq:E7components}, corresponding to the combination
\begin{equation}
\mathcal{U}_7 \equiv \Big[ E_{71} + E_{14} + E_{47} + E_0 \Big]_{s_{345} \rightarrow 0} \,.
\label{UfromE}  
\end{equation}
Note that from \eqn{eq:P7tree}, the coefficient of $E_0$ receives a $3/7$ contribution from $(71)$, and $2/7+2/7$ from $(14)$ and $(47)$.

Ignoring the tree amplitude, the quantity $\mathcal{U}_7$ is the exponential of $\log(A_7^{\rm NMHV}/A_7^{\rm BDS-like})$, whose prediction is given in \eqn{lnA7Lsymb}.  Using \eqn{afact} to compute $\mathcal{U}_7$ from \eqn{UfromE} in terms of the letters $a_{ij}$, we find at one, two, and three loops,
\begin{align}
\mathcal{U}_7^{(1)} &=
- \frac{1}{2} \log^2\bigg(\frac{a_{14}^2}{a_{11} a_{17}} \bigg)
- \frac{1}{2} \log^2 a_{11} - \frac{1}{2} \log^2 a_{17} \,,  \label{U71loop}\\
\mathcal{U}_7^{(2)} &= \frac{\Big(\mathcal{U}_7^{(1)} \Big)^2}{2!}
\,, \label{U72loops} \\ 
\mathcal{U}_7^{(3)} &= \frac{\Big(\mathcal{U}_7^{(1)} \Big)^3}{3!} \,.
\label{U73loops} 
\end{align}
Hence $\mathcal{U}_7$ exponentiates at symbol level, as predicted by \eqn{lnA7Lsymb}.  Substituting \eqn{a11tox} for $a_{11}$, and its cyclic permutations, into \eqn{U71loop}, we find perfect agreement with \eqn{lnA7Lsymb}.  We can also express the result in terms of the cross ratios $u_i$:
\be
\mathcal{U}_7^{(1)} = 
- \frac{1}{2} \log^2\bigg(\frac{u_1u_2}{u_3u_7}\bigg)
- \frac{1}{2} \log^2\bigg(\frac{u_1u_4u_5}{u_3u_6}\bigg)
- \frac{1}{2} \log^2\bigg(\frac{u_2u_6u_5}{u_7u_4}\bigg) \,.
\label{U71loopu}
\ee
Once this analysis is repeated at function level, we expect the first term in $\mathcal{U}_7^{(1)}$ to receive higher-loop zeta-valued contributions, dictated by the six-point function $U(x)$, while the last two terms simply get multiplied by the cusp anomalous dimension.

The last two terms in \eqn{U71loop} or \eqn{U71loopu} do not diverge in the factorization limit. On the other hand, they play an essential role in endowing $\mathcal{U}_7$ with the correct behavior as $p_7$ and $p_1$ become collinear. \Fig{Fig:Multifact7} shows that this collinear limit is well away from the factorization pole, in the sense of color ordering. So it should be possible to take this collinear limit ``inside'' the $s_{345}\to0$ multi-particle factorization limit, i.e.~as a further limit of it.

The $p_7 \parallel p_1$ collinear limit takes $x_{72}^2\to0$, and hence the cross ratio $u_5\to0$.  \Eqn{U71loopu} shows that the last two terms of $\mathcal{U}_7^{(1)}$ diverge logarithmically in this collinear limit, while the first term behaves smoothly.  Recall that the $n$-point BDS ansatz smoothly tends to the $(n-1)$-point BDS ansatz in all collinear limits.  However, this is not true for the BDS-like ansatz; that is, $Y_7 \not \rightarrow Y_6$ in collinear limits, rather it diverges logarithmically.  Essentially, the last two terms of \eqn{U71loop} account for this non-smooth behavior.  In the $p_7 \parallel p_1$ collinear limit,
\begin{align}
- \frac{1}{2} \log^2\bigg(\frac{a_{14}^2}{a_{11} a_{17}} \bigg)
&\xrightarrow{\ p_7 \parallel p_1 \ }
- \frac{1}{2} \log^2\bigg(\frac{u w}{v} \bigg) \,, \\
- \frac{1}{2} \log^2 a_{11} - \frac{1}{2} \log^2 a_{17} + Y_7
&\xrightarrow{\ p_7 \parallel p_1 \ } Y_6 \,.
\end{align}
Thus the last two terms in \eqn{U71loop} precisely account for the non-smooth collinear behavior of the BDS-like-normalized amplitude at seven points, within the multi-particle factorization limit.


\section{Discussion}
\label{Section:Discussion}

Following the inclusion of the Steinmann relations in the hexagon function bootstrap program~\cite{Caron-Huot:2016owq}, we have applied these constraints to heptagon symbols, in order to drastically reduce the number of symbols needed to bootstrap seven-point scattering amplitudes. We have been able to construct a basis of Steinmann heptagon symbols through weight 7, and those which further satisfy the MHV final-entry condition at weight 8. In order to apply the Steinmann relations transparently, we have shifted our focus from the familiar BDS-normalized amplitudes to BDS-like normalized analogues.  The simple conversions~\eqref{eq:BDSLikeToBDS} and~\eqref{eq:ERrelations} between functions in these two normalizations allow us to simultaneously take advantage of the smaller space of Steinmann heptagon symbols, and utilize the simple behavior exhibited by BDS-normalized functions near the collinear limit. With these advances, we have completely determined, in a conceptually simple manner, the symbols of the seven-point three-loop NMHV and four-loop MHV amplitudes in planar $\mathcal{N}=4$ SYM theory.

Calculating the symbol of these particular component amplitudes is only the tip of the Steinmann iceberg. The main limiting factor in applying the bootstrap at higher weight is the computational complexity resulting from the size of the space of Steinmann heptagon functions, which still grows close to exponentially, despite its small size relative to the general heptagon function space. This growth can be especially prohibitive when generating the general basis of Steinmann heptagon symbols at each higher weight. At the same time, nearly the entire space of Steinmann heptagon symbols is needed to describe the amplitudes we have bootstrapped -- including derivatives (coproducts) of higher-loop amplitudes. That is, the full space of Steinmann heptagon symbols is spanned by the derivatives of our amplitudes at weights 2 and 3. Only 15 of the 322 Steinmann heptagon symbols are absent from the span of these derivatives at weight 4. This situation resembles what is observed in the hexagon function bootstrap~\cite{Caron-Huot:2016owq}, where the derivatives of the five-loop six-point amplitude also span the full weight-2 and weight-3 Steinmann hexagon symbol spaces, while only 3, 12, and 30 symbols are absent from the span of these derivatives at weights 4, 5, and 6. In the hexagon case, all of these symbols are observed to drop out due to lower-weight restrictions on the appearance of zeta values (i.e.~the zeta values only appear in certain linear combinations with other hexagon functions, and this leads to symbol-level restrictions at higher weights).  We expect that a similar set of function-level restrictions will explain why a small set of weight-4 Steinmann heptagon symbols are not needed to describe the seven-point amplitude. (Only 386 of the 1030 weight-5 Steinmann heptagon symbols are currently needed to describe the four-loop MHV and three-loop NMHV amplitudes, but here we expect significantly more of these symbols to be needed to describe coproducts of yet higher-loop contributions.)  No physical explanation for the restrictions on the occurrence of zeta values at six points has yet been discerned, indicating that there remains some physics to be discovered.

More generally, the task of upgrading our symbol-level results to full functions will be left to future work. A full functional representation would be valuable for checking seven-point predictions in both the near-collinear~\cite{Sever:2011pc,Basso:2013vsa,Basso:2013aha,Basso:2014nra,Belitsky:2015efa,Basso:2015rta,Basso:2015uxa,Belitsky:2016vyq} and multi-Regge limits~\cite{Bartels:2008ce,Bartels:2008sc,Lipatov:2010ad,Fadin:2011we,Lipatov:2012gk,Dixon:2012yy, Bartels:2011ge,Bartels:2013jna,Bartels:2014jya,Bargheer:2016eyp, Broedel:2016kls,DelDuca:2016lad}.  An important problem is to generalize the all-loop results for six-point scattering in the multi-Regge limit~\cite{Basso:2014pla} to the seven-point case. The full functional form of the seven-point amplitude could assist the construction of an all-loop multi-Regge heptagon formula.

Bootstrapping amplitudes with eight or more external legs will require more than a simple extension of the heptagon bootstrap presented in this work. Both the hexagon and heptagon bootstrap approaches depend on the assumption that the weight-$2L$ generalized polylogarithms can be built from a finite symbol alphabet, corresponding to an appropriate set of cluster $\mathcal{A}$-coordinates. Going to $n=8$, we move into a cluster algebra with infinitely many $\mathcal{A}$-coordinates. It is expected that only a finite number of letters will appear at any finite loop order, but it is currently unknown how to characterize what sets may appear. In principle, this information ought to follow from a careful consideration of the Landau singularities of these amplitudes (see for example refs.~\cite{Dennen:2015bet,DPSSV} for recent related work). There is hope that patterns may emerge at currently accessible loop orders, which may provide insight into the letters appearing for $n>7$.

\acknowledgments

We have benefitted from stimulating discussions and correspondence with S.~Caron-Huot, M.~von Hippel, A.~von Manteuffel, R.~Schabinger, A.~Storjohann, and C.~Vergu. LD thanks the Kavli Institute for Theoretical Physics (National Science Foundation grant NSF PHY11-25915) and NORDITA for hospitality. AM thanks the Higgs Centre of the University of Edinburgh for hospitality. MS acknowledges C.-I.~Tan for having long emphasized the importance of the Steinmann relations and is grateful to NORDITA and to the CERN theory group for hospitality and support during the course of this work. This work has been supported by the US Department of Energy under contracts DE--AC02--76SF00515 (LD, AM and GP) and DE--SC0010010 (TH and MS). JMD acknowledges support from the ERC grant ERC-2014-CoG 648630 IQFT.


\appendix

\section{The BDS and BDS-like Ans\"atze}
\label{Appendix:BDSAnsatz}

The BDS ansatz~\cite{Bern:2005iz} for the $n$-particle MHV amplitude (with the Parke-Taylor tree amplitude scaled out) is given by
\begin{equation}
M_{n} \equiv
\frac{A_n}{A_n^{(0)}} = \text{exp} \left[ \sum_{L=1}^{\infty} a^{L} \left( f^{(L)}(\epsilon) \frac{1}{2} M_{n}^{(1)}(L\epsilon) + C^{(L)} \right) \right]
\end{equation}
with
\begin{equation}
\label{BDSf}
f^{(L)}(\epsilon) = f_{0}^{(L)} + \epsilon f_{1}^{(L)} + \epsilon^{2} f_{2}^{(L)},
\end{equation}
and where $\epsilon$ is the dimensional regularization parameter in
$D = 4-2\epsilon$.
Here $f_{0}^{(L)}$ is the planar cusp anomalous dimension with
\begin{equation}\label{eq:f0togk}
f_{0}^{(L)} = \frac{1}{4} \gamma_{K}^{(L)}=\frac{1}{2} \Gamma_{\text{cusp}}^{(L)}\,,
\end{equation}
according to the definition~\eqref{gamma_cusp}. However, note that in the above relation the superscript $L$ refers to coefficients in the expansion with respect to $a=2g^2$, and not $g^2$.

For $n=7$, the BDS ansatz takes the form
\begin{equation}
A_{7}^{\text{BDS}} = A_{7}^{\text{MHV}(0)} \text{exp} \left[ \sum_{L=1}^{\infty} a^{L} \left( f^{(L)}(\epsilon) \frac{1}{2} M_{7}^{(1)}(L\epsilon) + C^{(L)} \right) \right].
\end{equation}
Here we have explicitly factored out $1/2$ from the definition of $M_{7}^{(1)}(\epsilon)$ appearing in the original BDS paper.
The seven-particle one-loop MHV amplitude (again with the tree amplitude scaled out) appearing in the BDS ansatz is given by
\begin{equation}
M_{7}^{(1)}(\epsilon) = - \frac{1}{\epsilon^{2}} \sum_{i=1}^{7} \left( \frac{\mu^{2}}{-s_{i,i+1}} \right)^{\epsilon} + F_{7}^{(1)}(0)+\mathcal{O}(\epsilon)
\end{equation}
where
\begin{equation}
\label{Eq:F}
F_{7}^{(1)}(0) =\sum_{i=1}^{7} \left[ -\log\left( \frac{-s_{i,i+1}}{-s_{i,i+1,i+2}} \right)\log\left( \frac{-s_{i+1,i+2}}{-s_{i,i+1,i+2}} \right) + D_{7,i} + L_{7,i} +\frac{3}{2} \zeta_{2} \right]
\end{equation}
with
\begin{equation}
D_{7,i} = -\Li_{2} \left( 1{-}\frac{s_{i,i+1}\,s_{i{-}1,i,i+1,i+2}}{s_{i,i+1,i+2}\,s_{i{-}1,i,i+1}} \right)
\end{equation}
and
\begin{equation}
L_{7,i} = -\frac{1}{2} \log\left( \frac{-s_{i,i+1,i+2}}{-s_{i,i+1,i+2,i+3}} \right)\log\left( \frac{-s_{i+1,i+2,i+3}}{-s_{i{-}1,i,i+1,i+2}} \right).
\end{equation}
Notice that all of the dependence on the three-particle Mandelstam invariants is contained within $F_{7}^{(1)}(0)$, so we will focus on determining its dependence. We can replace the four-particle invariants with three-particle invariants in both $D_{7,i}$ and $L_{7,i}$. The two equations then become
\begin{equation}
D_{7,i} = -\Li_{2} \left(1{-}\frac{s_{i,i+1}s_{i+3,i+4,i+5}}{s_{i,i+1,i+2}s_{i{-}1,i,i+1}} \right) \text{, } L_{7,i} = -\frac{1}{2} \log\left( \frac{s_{i,i+1,i+2}}{s_{i+4,i+5,i+6}} \right)\log\left( \frac{s_{i+1,i+2,i+3}}{s_{i+3,i+4,i+5}} \right).
\end{equation}

At this point, it is convenient to switch to the $n=7$ dual conformal cross ratios $u_{i}$, defined in terms of the Mandelstam variables by
\begin{equation}
u_{i} = u_{i+1,i+4}=\frac{s_{i+2,i+3}\,s_{i+5,i+6,i+7}}{s_{i+1,i+2,i+3}\,s_{i+2,i+3,i+4}}\,,
\end{equation}
where all indices are understood mod 7. We can see from this definition that $D_{7,i}$ can be expressed simply in the $u_{i}$ variables as $D_{7,i} = -\Li_{2} \left(1{-}u_{i{-}2} \right)$. Using the dilogarithm identity $\Li_{2}(z) + \Li_{2}(1{-}1/z) = -\frac{1}{2}\log^{2} z$, we then rewrite $D_{7,i} = \Li_{2} \left(1{-}1/u_{i{-}2} \right) + \frac{1}{2} \log^{2} u_{i{-}2} $, and express $F_{7}^{(1)}(0)$ as
\begin{align}
\begin{split}
F_{7}^{(1)}(0) = \sum_{i=1}^{7} & \left[ -\log\left( \frac{s_{i,i+1}}{s_{i,i+1,i+2}} \right)\log\left( \frac{s_{i+1,i+2}}{s_{i,i+1,i+2}} \right) + \Li_{2} \left(1{-}1/u_{i} \right)+ \frac{1}{2} \log^{2} u_{i} \right. \\
&\text{ } \left. -\frac{1}{2} \log\left( \frac{s_{i,i+1,i+2}}{s_{i+4,i+5,i+6}} \right)\log\left( \frac{s_{i+1,i+2,i+3}}{s_{i+3,i+4,i+5}} \right) + \frac{3}{2} \zeta_{2} \right].
\end{split}
\end{align}
After some algebra, $F_{7}^{(1)}(0)$ can be shown to be
\bea
F_{7}^{(1)}(0) &=& \sum_{i=1}^{7} \biggl[ \Li_{2} \left(1{-}\frac{1}{u_{i}} \right)
  + \frac{1}{2} \log\left(\frac{u_{i+2}u_{i{-}2}}{u_{i+3}u_{i}u_{i{-}3}}\right)
  \log u_{i} \nonumber\\
&&\hskip1cm\null
  + \log s_{i,i+1} \log\left( \frac{s_{i,i+1}s_{i+3,i+4}}{s_{i+1,i+2}s_{i+2,i+3}} \right)
  + \frac{3}{2} \zeta_{2} \biggr]. \label{Eq:F2}
\eea
In this form, we have conveniently isolated all of the three-particle invariants in the first two terms.

Now we would like to factor out the three-particle invariants from $F_{7}^{(1)}(0)$ because this removes their dependence from $M_{7}^{(1)}$ as well. We define the function
\begin{equation}
Y_7 = -\sum_{i=1}^7 \biggl[ \Li_2\left(1-\frac{1}{u_i}\right)
  + \frac{1}{2} \log \left(\frac{u_{i+2}u_{i{-}2}}{u_{i+3}u_{i}u_{i{-}3}}\right)
               \log u_i \biggr]
\end{equation}
so that adding the term $Y_{7}$ removes the three-particle invariants from $M_{7}^{(1)}$:
\bea
\hat{M}_7^{(1)}(\epsilon) &\equiv& M_7^{(1)}(\epsilon) + Y_7
\nonumber\\
&=& \sum_{i=1}^7 \biggl[
  - \frac{1}{\epsilon^{2}} \left( \frac{\mu^{2}}{-s_{i,i+1}} \right)^{\epsilon}
  + \log s_{i,i+1}
  \log\left( \frac{s_{i,i+1} \, s_{i+3,i+4}}{s_{i+1,i+2} \, s_{i+2,i+3}} \right)
  + \frac{3}{2} \zeta_{2} \biggr] \,.
\label{hatM7}
\eea
The BDS-like ansatz is defined to be the BDS ansatz with $M_{7}^{(1)}$ replaced by with $\hat{M}_{7}^{(1)}$, which does not depend on any three-particle invariant:
\begin{equation}
A_{7}^{\text{BDS-like}} = A_{7}^{\text{MHV}(0)} \, \text{exp} \left[ \sum_{L=1}^{\infty} a^{L} \left( f^{(L)}(\epsilon) \frac{1}{2} \left(M_{7}^{(1)}(L\epsilon) + Y_{7}\right) + C^{(L)} \right) \right],
\label{A7BDSlike}
\end{equation}
Factoring out the BDS ansatz explicitly, we have
\begin{equation}
A_{7}^{\text{BDS-like}} = A_{7}^{\text{BDS}} \, \text{exp} \left[ \sum_{L=1}^{\infty} \frac{a^{L}}{2} \left( f^{(L)}(\epsilon) Y_{7}\right) \right].
\end{equation}
Recall that in the BDS ansatz formulation, the limit $\epsilon \rightarrow 0$ is taken. Since $Y_{7}$ is independent of $\epsilon$, we can set $\epsilon \rightarrow 0$ in \eqn{BDSf} and rewrite the BDS-like ansatz as simply
\begin{equation}
A_{7}^{\text{BDS-like}} =A_{7}^{\text{BDS}} \, \text{exp} \left[\frac{Y_{7}}{4} \sum_{L=1}^{\infty} a^{L} \Gamma_{\text{cusp}}^{(L)} \right],
\end{equation}
where we have used the definition~\eqref{eq:f0togk}. After introducing $\Gamma_{\text{cusp}} = \sum_{L=1}^{\infty} a^{L} \Gamma_{\text{cusp}}^{(L)}$, defined in eq.~(\ref{gamma_cusp}), we finally arrive at a simple representation of the BDS-like ansatz as a function of the BDS ansatz, the cusp anomalous dimension $\Gamma_{\text{cusp}}$, and ${Y}_{7}$,
\begin{align}
A_{7}^{\text{BDS-like}} = A_{7}^{\text{BDS}} \, \text{exp} \left[\frac{ \Gamma_{\text{cusp}}}{4} Y_{7} \right].
\end{align}
This result can be generalized to any $n$ for which a suitable BDS-like ansatz exists, see \eqn{BDSlikeBDSn}.


\section{A Matrix Approach For Computing Integrable Symbols}
\label{Appendix:BootstrapApproach}

We provide here a conceptually simple method for generating a basis of integrable symbols, given the set of symbol letters on which they depend. This algorithm is iterative, and assumes that one has seeded the algorithm with a basis at low weight. For general heptagon symbols, this seed is provided at weight 1 by the first entry condition reviewed in section~\ref{Section:SymbolSingularityStructure}. It consists of the 7 weight-1 symbols corresponding to $\log a_{1i}$. For Steinmann heptagon symbols, the seed is provided by the 28 weight-2 heptagon symbols of the functions shown in eq.~(\ref{SteinmannheptWt2}).

Let $B^{(k)}$ denote a basis of symbols at weight $k$, and let $b_k = \dim B^{(k)}$. Let us also denote the $i$-th element of $B^{(k)}$ by $B^{(k)}_i$. Given $B^{(k)}$, we can make an ansatz for symbols of weight $(k{+}1)$ of the form
\begin{equation}
\label{eq:ansatz}
\sum_{i=1}^{b_k} \sum_{q=1}^{|\Phi|} c_{iq} \ B^{(k)}_i \otimes \phi_q\,,
\end{equation}
where the sum over $q$ runs over all letters in the symbol alphabet $\Phi$, i.e.~$\phi_q\in\Phi$, and the $c_{iq}$ are undetermined rational coefficients. The number of letters is denoted by $|\Phi|$.  The quantity~(\ref{eq:ansatz}) will be the symbol of some weight-$(k{+}1)$ function only if it satisfies the integrability constraints of~\eqn{eq:integrability} for all $j$. By construction, these constraints are automatically satisfied for $j=1,2,\ldots,k-1$, because the elements of $B^{(k)}$ are already valid, integrable symbols. It therefore remains only to impose integrability in the final two entries at weight $(k{+}1)$, i.e.~for $j=k$.

Each $B^{(k)}_i$ can of course be expressed as
\begin{equation}\label{eq:cfb}
B^{(k)}_i = \sum_{j = 1}^{b_{k-1}} \sum_{p=1}^{|\Phi|} f_{ijp} \ B^{(k-1)}_j \otimes \phi_p
\end{equation}
for some known coefficients $f_{ijp}$, so we can rewrite our ansatz as
\begin{equation}
\sum_{i=1}^{b_k} \sum_{j=1}^{b_{k-1}} \sum_{p,q=1}^{|\Phi|} c_{iq} f_{ijp} \ B_j^{(k-1)} \otimes \phi_p \otimes \phi_q\,.
\end{equation}
Denoting
\be\label{eq:Fpq}
F_{pq}=\sum_{i=1}^{b_k} \sum_{j=1}^{b_{k-1}}  c_{iq} f_{ijp} \ B_j^{(k-1)}\,,
\ee
the quantity \eqref{eq:cfb} satisfies integrability in the final two entries only if
\begin{equation}\label{eq:k11integrability}
\sum_{p,q=1}^{|\Phi|} F_{pq}\, d \log \phi_p \wedge d \log \phi_q = 0\,,
\end{equation}
where the wedge product between two letters $\phi_p,\phi_q$ that are functions of the independent variables $x^i$ is defined as
\begin{equation}
d\log\phi_p\wedge d\log\phi_q=\sum_{m,n}\left[\frac{\partial \log\phi_p}{\partial x^m}\frac{\partial \log\phi_q}{\partial x^n}-\frac{\partial \log\phi_p}{\partial x^n}\frac{\partial \log\phi_q}{\partial x^m}\right]dx^m \wedge dx^n\,.
\end{equation}
The term in brackets above will be a rational function of the independent variables, which can be turned polynomial by multiplying with the common denominator, without altering the equations \eqref{eq:k11integrability}. Each independent polynomial factor of the $x_i$ times their differentials must vanish separately, which leads to distinct rational equations for the $F_{pq}$. If the number of linearly independent equations is $r$, then we may equivalently write eq.~\eqref{eq:k11integrability} as
\begin{equation}
\sum_{p,q=1}^{|\Phi|} F_{pq} W_{pql} = 0\,,\quad \forall l \in \{1,2,\ldots,r\}\,,
\end{equation}
in terms of a rational tensor $W_{pql}$. Taking the tensor product of the indices $p,q$ we may think of $W$ as a $|\Phi|^2\times r$ matrix, or rather a ${{|\Phi|}\choose{2}} \times r$ matrix after taking into account its antisymmetry in $p\leftrightarrow q$.

Since the $B_j^{(k-1)}$ are elements of the basis $B^{(k-1)}$ of weight-$(k{-}1)$ symbols, they are linearly independent. Each term in the sum over $j$ in \eqref{eq:Fpq} must therefore vanish separately. In this manner, we finally arrive at the following set of $r\times b_{k-1}$ linear constraints on the $|\Phi|\times b_k$ unknown coefficients $c_{iq}$:
\begin{equation}
\sum_{i=1}^{b_k} \sum_{p,q} c_{iq} f_{ijp} W_{pql} = 0\,, \qquad \forall j \in \{1,2,\ldots,b_{k-1}\}\,,\quad  l \in \{1,2,\ldots,r\}\,.
\label{eq:constraints}
\end{equation}

We now specialize to the case of interest in this paper by adopting the 42-letter symbol alphabet presented in eqs.~\eqref{Salphabet} and~\eqref{Salphabetpermutation}. There are 132 vanishing linear combinations of the 861 objects $d \log \phi_p \wedge d \log \phi_q$, i.e.~there are 132 irreducible weight-2 integrable symbols (these are in correspondence with elements of the so-called Bloch group $B_2$; see for example ref.~\cite{Golden:2013xva}). This means that there are $r = 861 - 132 = 729$ nontrivial integrability constraints for the heptagon symbol alphabet. In solving the linear constraints~(\ref{eq:constraints}) for the $c_{iq}$, we are free to replace $W$ by any matrix which spans the same image as $W$ without changing the content of the constraints. It is highly advantageous to choose a basis for the image of $W$ that is as sparse as possible, and which has numerical entries as simple as possible. In our bootstrap we used a representation of the image of $W$ as a $861 \times 729$ matrix\footnote{To orient the reader already familiar with the hexagon bootstrap: there the symbol alphabet has size $|\Phi| = 9$, and there are 10 irreducible weight-2 integrable symbols, so the $W$ matrix for the hexagon alphabet has size $36 \times 26$.} with only 1195 nonzero entries having values $\pm 1$.

Finally, then, the integrability constraints shown in eq.~(\ref{eq:constraints}) take the form of $729\, b_{k-1}$ linear equations on the $42\, b_k$ unknowns $c_{iq}$. Finding a basis for the nullspace of this $729\, b_{k-1} \times 42\, b_k$ linear system provides a basis for $B^{(k+1)}$, the integrable symbols at weight $k+1$. For the purposes of the Steinmann heptagon bootstrap, we have further cut down the weight-2 basis yielded by this procedure to only those 28 symbols that satisfy the Steinmann relations before proceeding to weight 3. We have carried out the large linear algebra problems necessary for the heptagon bootstrap with the help of the {\tt SageMath} system~\cite{SageMath}, which employs the {\tt IML} integer matrix library~\cite{Chen:2005:BBC:1073884.1073899}. As a double check, we also fed the weight-7 integrability constraint matrix into A.~von Manteuffel's {\tt FinRed} program, which independently generated a basis for the 9570-dimensional weight-7 Steinmann heptagon space reported in table~\ref{tab:t1}.

\bibliographystyle{JHEP}
\bibliography{heptagon}

\end{document}